\documentclass[12pt]{article}

\usepackage{arxiv}

\usepackage{blindtext} % for dummy text
\usepackage{graphicx} % figures etc.
\usepackage[utf8]{inputenc} % allow utf-8 input
\usepackage[T1]{fontenc}    % use 8-bit T1 fonts
\usepackage[hidelinks]{hyperref}       % hyperlinks
\usepackage{url}            % simple URL typesetting
\usepackage{booktabs}       % professional-quality tables
\usepackage{amsmath,amsfonts,amssymb}       % blackboard math symbols
\usepackage{nicefrac}       % compact symbols for 1/2, etc.
\usepackage[final]{microtype}      % microtypography
\usepackage{xr} % for cross-referencing with external documens (SI)
\usepackage{etoolbox,siunitx} %for the \num command, nice formatting for numbers
\usepackage{booktabs} %better tables (\toprule \bottomrule \midrule \cmidrule)
\usepackage{multirow} %allows fancier tables
\usepackage{color}
\usepackage[width=.75\textwidth]{caption}

\setcitestyle{citesep={,}}

\makeatletter
\DeclareRobustCommand\citenum
{\begingroup
	\NAT@swatrue\let\NAT@ctype\z@\NAT@parfalse\let\textsuperscript\relax% NEW
	\NAT@citexnum[][]}
\makeatother

\newcommand{\nn}{PhysNet}

\title{PhysNet: A Neural Network for Predicting Energies, Forces,
  Dipole Moments and Partial Charges}

\author{
  Oliver T.~Unke\footnotemark[1]\\
  Department of Chemistry\\
  University of Basel\\
  Basel, Switzerland \\
  \texttt{oliver.unke@unibas.ch} \\
   \And
 Markus Meuwly\thanks{to whom correspondence should be addressed}\\
 Department of Chemistry\\
 University of Basel\\
 Basel, Switzerland \\
  \texttt{m.meuwly@unibas.ch} \\
}

\newcommand{\boldentry}[2]{%
	\multicolumn{1}{S[table-format=#1,
		mode=text,
		text-rm=\fontseries{b}\selectfont
		]}{#2}}

\begin{document}
\maketitle

\begin{abstract}
In recent years, machine learning (ML) methods have become
increasingly popular in computational chemistry. After being trained on
appropriate \textit{ab initio} reference data, these methods allow to
accurately predict the properties of chemical systems, circumventing
the need for explicitly solving the electronic Schr\"odinger
equation. Because of their computational efficiency and scalability to
large datasets, deep neural networks (DNNs) are a particularly
promising ML algorithm for chemical applications.
This work introduces \nn, a DNN architecture designed for predicting
energies, forces and dipole moments of chemical systems. \nn\ achieves
state-of-the-art performance on the QM9, MD17 and ISO17
benchmarks. Further, two new datasets are generated in order to probe
the performance of ML models for describing chemical reactions,
long-range interactions, and condensed phase systems. It is shown that
explicitly including electrostatics in energy predictions is crucial
for a qualitatively correct description of the asymptotic regions of a
potential energy surface (PES).
\nn\ models trained on a systematically constructed set of small
peptide fragments (at most eight heavy atoms) are able to generalize
to considerably larger proteins like deca-alanine (Ala$_{10}$): The
optimized geometry of helical Ala$_{10}$ predicted by \nn\ is
virtually identical to \textit{ab initio} results
($\mathrm{RMSD}=0.21$~\AA). By running unbiased molecular dynamics
(MD) simulations of Ala$_{10}$ on the \nn-PES in gas phase, it is
found that instead of a helical structure, Ala$_{10}$ folds into a
``wreath-shaped'' configuration, which is more stable than the helical
form by $0.46$~kcal~mol$^{-1}$ according to the reference \textit{ab
  initio} calculations.

\end{abstract}

\keywords{\nn \and  Computational chemistry \and Neural network \and Machine learning \and Potential energy surface \and Molecular dynamics}

% % % % % % % % % % % % % % % % % % % % % % % % % % % % % % %
% INTRODUCTION
% % % % % % % % % % % % % % % % % % % % % % % % % % % % % % %
\section{Introduction}
\label{sec:introduction}
As was stated by Dirac already in 1929,\cite{dirac1929quantum} the
Schr\"odinger equation (SE) in principle contains all that is
necessary to describe the whole of chemistry. Unfortunately, the SE
can only be solved in closed form for the simplest systems, hence
computational and numerical methods have been devised to find
approximate solutions. However, even with these approximations,
solving the electronic SE is computationally demanding and, depending
on the accuracy required and the approximations used, is only
tractable for a limited number of atoms.\cite{pople1999quantum}

For this reason, machine learning (ML) methods have become
increasingly popular in recent years in order to circumvent the
solution of the SE altogether. Such approaches give a computer the
ability to learn patterns in data without being explicitly
programmed\cite{samuel2000some} and have been used in the past to
estimate properties of unknown compounds or structures after being
trained on a reference
dataset.\cite{rupp2012fast,montavon2013machine,hansen2013assessment,hansen2015machine}
Of particular interest in this context is the energy $E$, as the
forces derived from it can be used to drive molecular dynamics (MD)
simulations. Since the Hamiltonian of a chemical system is uniquely
determined by the external potential, which in turn depends on a set
of nuclear charges $\{Z_i\}$ and atomic positions $\{\mathbf{r}_i\}$,
all information necessary to determine $E$ is contained in $\{Z_i,
\mathbf{r}_i\}$. Hence, there must exist an exact mapping $f:\{Z_i,
\mathbf{r}_i\} \mapsto E$, which is usually referred to as a potential
energy surface (PES).

Artificial neural networks
(NNs)\cite{mcculloch1943logical,kohonen1988introduction,abdi1994neural,bishop1995neural,clark1999neural,ripley2007pattern,haykin2009neural}
are a particular class of ML algorithms proven to be general function
approximators\cite{gybenko1989approximation,hornik1991approximation}
and thus ideally suited to learn a representation of the PES. For
small systems, PESs based on NNs have been designed in the spirit of a
many-body
expansion,\cite{manzhos2006random,manzhos2007using,malshe2009development}
but these approaches scale poorly for large systems, because they
typically involve a large number of individual NNs (one for each term
in the many-body expansion). An alternative approach, known as
high-dimensional neural network (HDNN),\cite{behler2007generalized}
decomposes the total energy of a chemical system into atomic
contributions and uses a single NN (or one for each element) for the
energy prediction. It relies on the chemically intuitive assumption
that the contribution of an atom to the total energy depends mainly on
its local environment.

Two variants of HDNNs can be distinguished: One is
``descriptor-based'', also referred to as Behler-Parrinello
networks,\cite{behler2007generalized} for which the environment of an
atom is encoded in a hand-crafted
descriptor,\cite{behler2011atom,khorshidi2016amp,artrith2017efficient,unke2018reactive}
which is used as input of a multi-layer feed-forward NN. Examples for
this kind of approach are ANI\cite{smith2017ani} and
TensorMol.\cite{yao2018tensormol} In the second
``message-passing''\cite{gilmer2017neural} variant, nuclear charges
and Cartesian coordinates are used as input and a deep neural network
(DNN) is used to exchange information between individual atoms, such
that a meaningful representation of their chemical environments is
learned directly from the data. This approach was first introduced by
the DTNN\cite{schutt2017quantum} and has since been refined in other
DNN architectures, for example SchNet\cite{schutt2017schnet} or
HIP-NN.\cite{lubbers2018hierarchical} While both types of HDNN perform
well, the message-passing variant was found to be able to adapt better
to the training data and usually achieves a better
performance.\cite{schutt2018quantum}

This work introduces \nn, a HDNN of the message-passing type designed
based on physical principles. It is shown that \nn\ improves upon or
matches state-of-the-art performance on the
QM9,\cite{ramakrishnan2014quantum} MD17,\cite{chmiela2017machine} and
ISO17\cite{schutt2017schnet} benchmark data sets. Further, two new
benchmark datasets are presented: The first set contains structures
probing the PES of S$_{\rm N}$2 reactions of methyl halides with
halide anions. Capturing the correct long-range physics is
particularly challenging for this dataset due to the presence of
strong charge-dipole interactions. By comparing different variants of
\nn, it is demonstrated that the explicit inclusion of electrostatics
in the energy prediction is important for correctly describing
asymptotic regions of the PES.

The second dataset contains structures for small protein-like
compounds (consisting of at most eight heavy atoms) and clusters with
water molecules. Available benchmark datasets cover conformational and
chemical degrees of freedom, but they usually do not probe many-body
intermolecular interactions, which are important in the description of
condensed phase systems. Due to their biological importance, proteins
in aqueous solution are a particularly relevant system of this
kind. However, even small proteins contain hundreds of atoms, which
makes \textit{ab initio} reference calculations for them prohibitively
expensive. Fortunately, it is possible to construct a predictive ML
model for large molecules by training it only on smaller molecules
that are structurally similar.\cite{huang2017chemical} These so-called
``amons'' can be readily constructed by considering a large molecule
as chemical graph, generating all possible connected subgraphs with a
fixed number of heavy atoms and saturating the resulting structures
with hydrogen atoms.\cite{huang2017chemical} Due to the fact that most
proteins are comprised of only \num{20} different amino acids, many
bonding patterns are shared between proteins and a comparatively small
number of amons is sufficient to cover all possibilities.

It is shown that a \nn\ model trained on this data is able to
accurately predict interaction energies of common sidechain-sidechain
and backbone-backbone interactions in proteins. The model also
generalizes to considerably larger molecules than it was trained on:
When an ensemble\cite{breiman1996bagging} of \nn\ models is used to
optimize the geometry of helical deca-alanine (Ala$_{10}$), the
results are almost indistinguishable from a structure optimized at the
reference DFT level of theory ($\mathrm{RMSD} = 0.21$~\AA). By running
unbiased MD simulations on the \nn-PES, a ``wreath-shaped''
configuration of Ala$_{10}$ is found, which, according to \textit{ab
  initio} calculations, is as stable as the helical form in gas phase
(the wreath-shaped form is even slightly lower in energy by
$0.46$~kcal~mol$^{-1}$).

In section~\ref{sec:methods}, the \nn\ architecture and the process
used for generating the reference data for S$_{\rm N}$2 reactions and
solvated protein fragments is described in detail. The performance of
\nn\ on these datasets and other commonly used quantum chemical
benchmarks is reported in section~\ref{sec:results}. Further, the
generalization capabilities of the model are explored by applying it
to Ala$_{10}$. Finally, the results are discussed and summarized in
section~\ref{sec:discussion_and_conclusion}.

% % % % % % % % % % % % % % % % % % % % % % % % % % % % % % %
% METHODS
% % % % % % % % % % % % % % % % % % % % % % % % % % % % % % %
\section{Methods}
\label{sec:methods}
In section \ref{sec:neural_network}, the building blocks of \nn\ and
its complete architecture are described. Further, the procedure used
for training \nn\ on reference data (how the neural network parameters
are fitted) is described in section \ref{sec:training}. Because
currently available benchmark datasets do not cover chemical reactions or many-body
intermolecular interactions (as exhibited by condensed phase systems), two additional sets of reference data for
S$_{\rm N}$2 reactions and solvated protein fragments were constructed
and their generation is detailed in
section~\ref{sec:dataset_generation}.

% NEURAL NETWORK
\subsection{Neural network}
\label{sec:neural_network}

The basic building block of every fully-connected NN are so-called
dense layers. They take an input vector $\mathbf{x} \in
\mathbb{R}^{n_{\rm in}}$ and output a vector $\mathbf{y} \in
\mathbb{R}^{n_{\rm out}}$ based on the transformation
\begin{equation}
\mathbf{y}= \mathbf{W}\mathbf{x} +\mathbf{b}
\label{eq:dense_layer}
\end{equation}
where $\mathbf{W} \in \mathbb{R}^{n_{\rm out}\times n_{\rm in}}$ and
$\mathbf{b}\in \mathbb{R}^{n_{\rm out}}$ are parameters and $n_{\rm
  in}$ and $n_{\rm out}$ denote the dimensionality of input and output
vectors, respectively. Note that a single dense layer can only
represent a linear transformation from input to output. In order to
model arbitrary non-linear relationships, at least two dense layers
need to be stacked and combined with a (non-linear) activation
function $\sigma$, i. e.
\begin{equation}
\mathbf{y} = \mathbf{W_2}\sigma\left(\mathbf{W_1}\mathbf{x} +\mathbf{b_1}\right) + \mathbf{b_2}
\label{eq:shallow_neural_network}
\end{equation}
Throughout this work, the notation $\sigma(\mathbf{x})$ means that a
scalar function $\sigma$ is applied to a vector $\mathbf{x}$
entrywise, i.e.\ $\sigma(\mathbf{x}) =
[\sigma(x_1)\ \cdots\ \sigma(x_{n_{\rm in}})]^{\mathsf{T}}$.

\begin{figure}[htbp]
\centering
\includegraphics[width=0.75\textwidth]{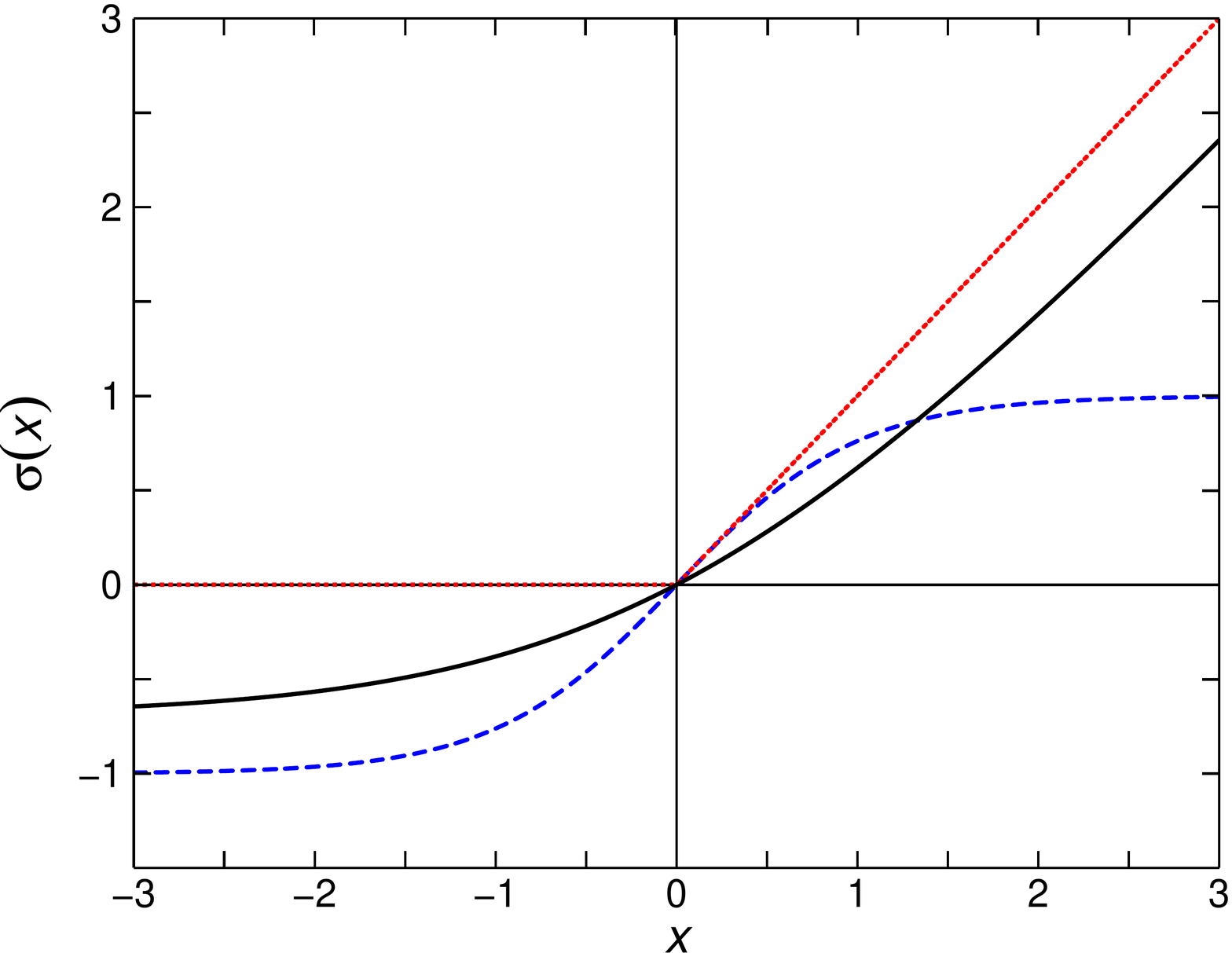}
\caption{Popular activation functions $\sigma(x)$. The solid black
  line shows the shifted softplus activation used in this work, the
  dotted black line shows $\max\left(0,x\right)$ and the dashed blue
  line shows $\tanh(x)$. Note that $\max\left(0,x\right)$ is not
  differentiable at $x=0$, which causes problems when continuous derivatives of
  the NN output are of interest, and $\tanh(x)$ saturates for large
  $\lvert x\rvert$, which makes training deep neural networks
  difficult.}
\label{fig:activation_function}
\end{figure}

Two dense layers combined according to
Eq.~\ref{eq:shallow_neural_network} can already approximate any
(non-linear) mapping between input $\mathbf{x}$ and output
$\mathbf{y}$, provided that the first (``hidden'') layer is ``wide''
enough (contains sufficiently many neurons) and an appropriate
activation function $\sigma$ is
used.\cite{gybenko1989approximation} Many different choices for
$\sigma$ are possible,\cite{chen1995universal} for example, popular
activation functions include $\tanh(x)$\cite{gybenko1989approximation}
or $\max\left(0,x\right)$.\cite{hahnloser2000digital} In this work,
the shifted softplus function\cite{schutt2017schnet} given by
$\sigma(x) = \log\left(e^x + 1\right) - \log\left(2\right)$ is used
(see Fig.~\ref{fig:activation_function}).

While \textit{shallow} neural networks composed of just two dense
layers (Eq.~\ref{eq:shallow_neural_network}) are already capable of
approximating arbitrary functions,\cite{chen1995universal}
\textit{deep} neural networks composed of more than two layers were
shown to be exponentially more parameter efficient in their
approximation capability.\cite{eldan2016power}

In this work, several reusable building blocks are combined in a
modular fashion to form a deep neural network architecture, which
predicts atomic contributions to properties (such as energy) of a
chemical system composed of $N$ atoms based on atomic features
$\mathbf{x}_{i} \in \mathbb{R}^F$ (here, $F$ denotes the
dimensionality of the feature space). The features simultaneously
encode information about nuclear charge $Z$ and local atomic
environment of each atom $i$ and are constructed by iteratively
refining an initial representation depending solely on $Z_i$ through
coupling with the feature vectors $\mathbf{x}_{j}$ of all atoms $j\neq
i$ within a cut-off radius $r_{\rm cut}$.

\begin{figure}[htbp]
\captionsetup{width=1.0\textwidth}
\centering
\includegraphics[width=1.0\textwidth]{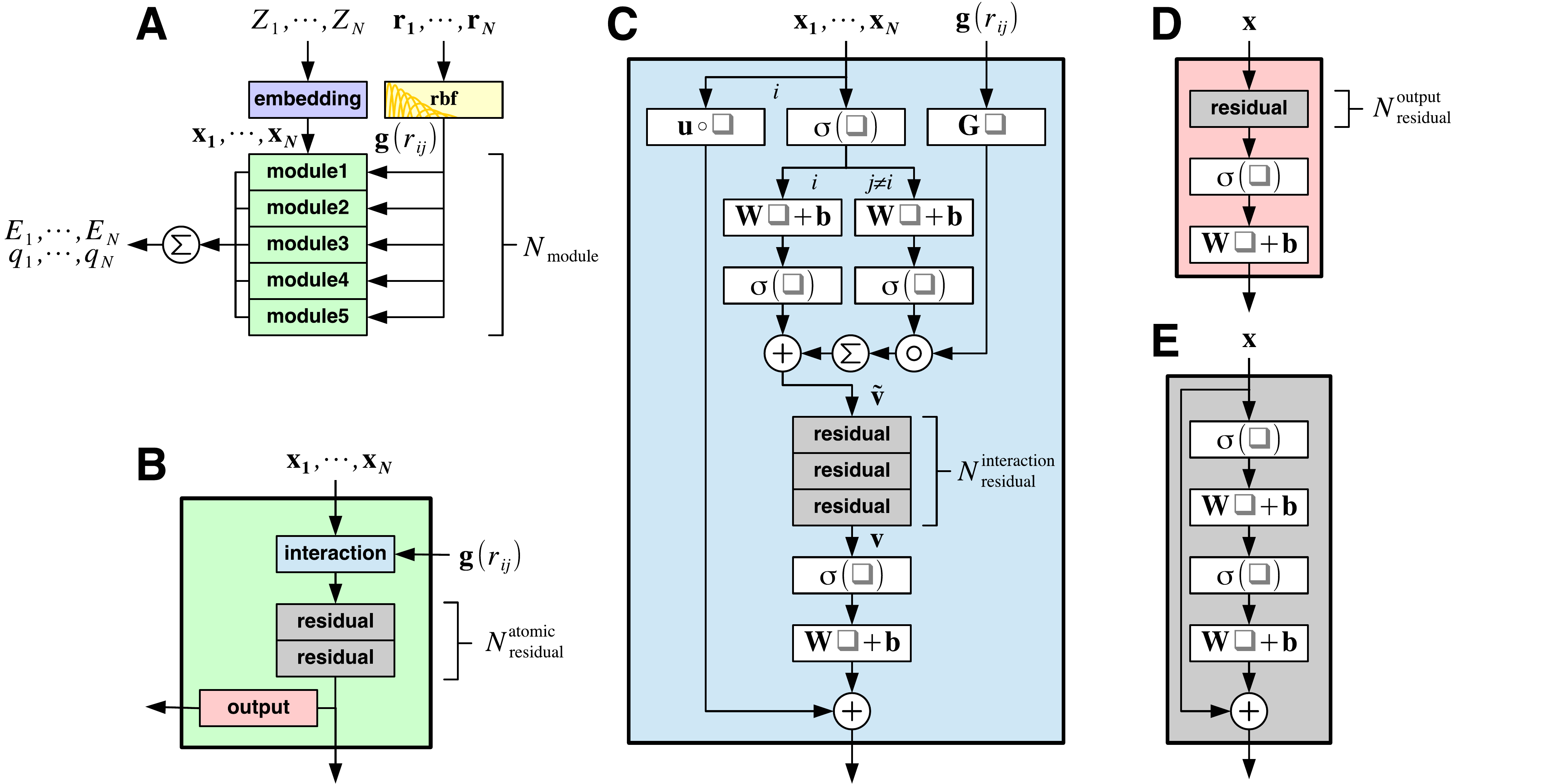}
\hspace{1cm}
\caption{\textbf{A}: Overview over the \nn\ architecture. The input
  nuclear charges $Z_i$ of $N$ atoms are transformed to feature
  vectors $\mathbf{x}_i\in \mathbb{R}^F$ via an embedding layer
  (purple, Eq.~\ref{eq:embedding_layer}) and passed iteratively
  through a stack of $N_{\rm module}$ modular building blocks
  (green). From the input Cartesian coordinates $\mathbf{r}_i$, all
  pairwise distances within a cut-off radius $r_{\rm cut}$ are
  calculated and expanded in a set of $K$ radial basis functions (rbf,
  yellow, Eq.~\ref{eq:radial_basis_function}) forming the entries of
  the vectors $\mathbf{g}(r_{ij}) \in \mathbb{R}^K$, which are
  additional inputs to each module. The output of all modules is
  summed to form the final atom-wise predictions of the neural
  network, e.g.\ atomic energy contributions $E_i$ and partial charges
  $q_i$ (Eq.~\ref{eq:final_prediction}). \textbf{B}: Structure of a
  modular building block. Each module transforms its input through an
  interaction block (blue) followed by $N_{\rm residual}^{\rm atomic}$
  residual blocks (grey). The computation then splits into two
  branches: One branch transforms the input further through an output
  block (red) to form the module output, whereas the other branch
  passes the transformed input directly to the next module in the
  hierarchy. \textbf{C}: Interaction block. After passing through the
  activation function $\sigma$, the incoming features of the
  central atom $i$ and neighbouring atoms $j$ split paths and are
  further refined through separate non-linear dense layers. The
  attention mask $\mathbf{G}\mathbf{g}(r_{ij})$ selects features of
  atoms $j$ based on their distance to atom $i$ and adds them to its
  features in order to compute the proto-message $\mathbf{\tilde{v}}$
  (Eq.~\ref{eq:interaction_block_intermediate}), which is refined
  through $N_{\rm residual}^{\rm interaction}$ residual blocks (grey)
  to the message $\mathbf{v}$. After an additional activation and
  linear transformation, $\mathbf{v}$, which represents the
  interactions between atoms, is added to the gated feature
  representations $\mathbf{u}\circ\mathbf{x}$
  (Eq.~\ref{eq:interaction_block_final}). \textbf{D}: Output block. An
  output block passes its input through $N_{\rm residual}^{\rm
    output}$ residual blocks (grey) and a dense layer (with linear
  activation) to compute the final output of a module
  (Eq.~\ref{eq:output_block}). \textbf{E}: Pre-activation residual
  block. Each residual block refines its input by adding a residual
  computed by a two-layer neural network
  (Eq.~\ref{eq:residual_block}). However, note that the usual order of
  dense layers and activations is reversed, which allows unrestricted
  gradient flow when training the neural network.\cite{he2016identity} The values of the
  hyperparameters ($N_{\rm module}$, $N_{\rm residual}^{\rm atomic}$,
  $N_{\rm residual}^{\rm interaction}$, $N_{\rm residual}^{\rm
    output}$, \dots) used in this work are given in
  Table~\ref{tab:hyperparameters}.}
	\label{fig:nn_architecture}
\end{figure}

In the following, the individual building blocks of the neural network
and important underlying concepts are described in more detail. The
complete architecture is schematically represented in
Fig.~\ref{fig:nn_architecture}. Note that for the remainder of this
section, superscripts $l$ are used to denote features or parameters of
layer $l$. All models are implemented in the
TensorFlow\cite{tensorflow2015} framework.

\paragraph{Embedding layer}
An embedding is a mapping from a discrete object to a vector of real
numbers. For example, word embeddings\cite{mikolov2013distributed}
find wide spread use in the field of natural language processing,
where semantically similar words are mapped such that they appear
close to each other in the embedding space. In this work, atomic
numbers $Z \in \mathbb{N}$ are mapped to embeddings $\mathbf{e}_Z \in
\mathbb{R}^F$, where the entries of the embedding vectors
$\mathbf{e}_Z$ are parameters. The embedding layer initializes the
atomic features of an atom with nuclear charge $Z$ to the
corresponding embedding vector $\mathbf{e}_Z$
(Eq.~\ref{eq:embedding_layer}).
\begin{equation}
\mathbf{x}_{i}^{0} = \mathbf{e}_{Z_i}
\label{eq:embedding_layer}
\end{equation}
The output of the embedding layer is passed to a stack of $N_{\rm
  module}$ modules sharing the same composition, but not parameters.

\paragraph{Module}
Each module in the stack (except for the first one, which receives its
input from the embedding layer) takes the output of the previous
module and couples the features $\mathbf{x}_i$ of each atom $i$ with
the features $\mathbf{x}_j$ of all atoms $j$ within the cut-off
distance $r_{\rm cut}$ through an interaction block. The features are
then further refined atom-wise through $N^{\rm atomic}_{\rm residual}$
residual blocks. Subsequently, the computation splits into two
branches: One branch passes the atomic features onwards to the next
module in the stack (if present) without further modification, whereas
the other branch passes the features to an output block, which
computes the module's contribution to the final prediction. The split
into two branches helps to decouple the feature representations passed
between modules from the prediction task at hand.

\paragraph{Residual block}
The ability of a neural network to model arbitrary functions should
always increase, or at least remain the same, when the depth
(i.e.\ the amount of dense layers stacked on top of each other) is
increased, as additional layers could in principle always reduce to
the identity mapping and should therefore never decrease the
performance. However, this is not observed in practice: As neural
networks get deeper, they become increasingly difficult to train
because of the vanishing gradients
problem,\cite{glorot2010understanding} which leads to a degradation of
their performance. In order to alleviate this, it was proposed to add
shortcut connections to the neural network architecture that skip one
or several layers, creating a so-called residual
block.\cite{he2016deep}

Since their first introduction, the design of residual blocks was
further refined to allow completely unhindered gradient flow through
all layers of a neural network.\cite{he2016identity} It was shown that
stacking these so-called pre-activation residual blocks allows
successfully training neural networks more than \num{1000} layers
deep.\cite{he2016identity} In this work, pre-activation residual
blocks are used extensively to refine the atomic features according to
\begin{equation}
\mathbf{x}_{i}^{l+2} = \mathbf{x}_{i}^{l} +
\mathbf{W}^{l+1}\sigma\left(\mathbf{W}^{l}\sigma(\mathbf{x}_{i}^{l}) +
\mathbf{b}^{l}\right) + \mathbf{b}^{l+1}
\label{eq:residual_block}
\end{equation}
where $\mathbf{x}_{i}^{l}$ and $\mathbf{x}_{i}^{l+2}$ are input and
output features, respectively, and $\mathbf{W}^{l}, \mathbf{W}^{l+1}
\in \mathbb{R}^{F\times F}$ and $\mathbf{b}^{l}, \mathbf{b}^{l+1} \in
\mathbb{R}^{F}$ are parameters.

\paragraph{Interaction block}
In order to predict properties that depend on the environment of an
atom $i$, it is important to model interactions with its surrounding
atoms $j$ in a chemically and physically meaningful manner. In doing
so, it is crucial that all known invariances of the property of
interest are respected: For example, the energy of a molecular system
is known to be invariant with respect to translation, rotation and
permutation of equivalent atoms.\cite{behler2011atom} Further, since
it is a valid assumption that most (but not all) chemical interactions
are inherently short-ranged,\cite{unke2018reactive} it is meaningful
to introduce a cut-off radius $r_{cut}$, such that only interactions
within the local environment of an atom are considered. Apart from
encoding chemical knowledge directly in the modelling of interactions,
this approach has the important computational advantage of making
predictions scale linearly with system size.

\begin{figure}[htbp]
\centering
\includegraphics[width=0.75\textwidth]{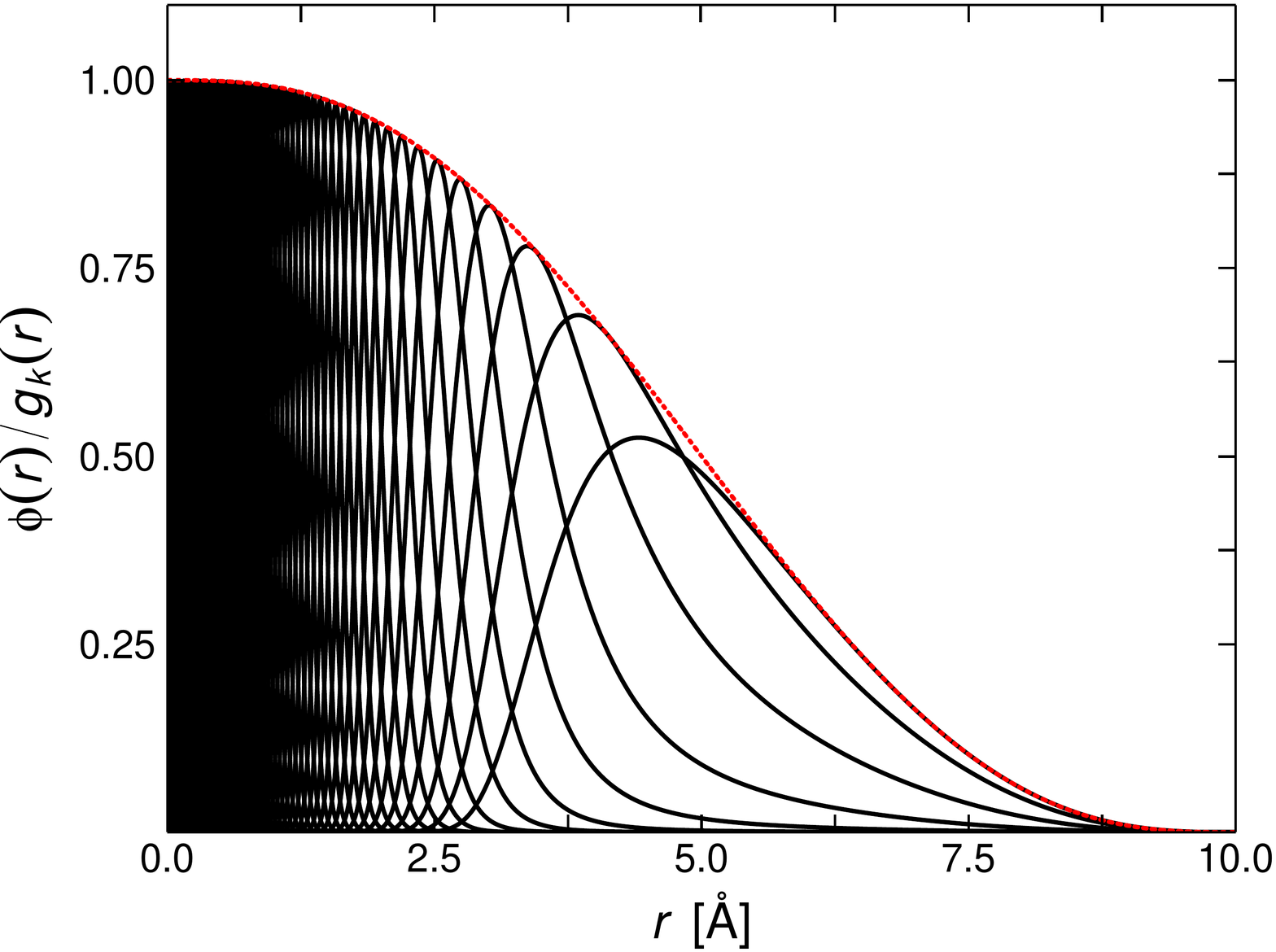}
\caption{Radial basis functions. Shown are the cutoff-function
  $\phi(r)$ (dotted red curve, Eq.~\ref{eq:cut-off_function}) along
  with $K=\num{64}$ radial basis functions $g_k(r)$ (solid black
  curves, Eq.~\ref{eq:radial_basis_function}) with a fixed $\beta_k =
  \left(2K^{-1}\left(1-\exp(-r_{\rm cut})\right)\right)^{-2}$ and
  values of $\mu_k$ equally spaced between $\exp(-r_{\rm cut})$ and
  $1$ with $r_{\rm cut}=10$ \AA. Note that for larger $r$, the basis
  functions automatically become wider even though all $g_k(r)$ share
  the same width parameter $\beta_k$.}
\label{fig:radial_basis_functions}
\end{figure}

Based on these design principles, the feature vector $\mathbf{x}$ of
an atom is refined by interacting with its local environment through a
``message''\cite{gilmer2017neural} $\mathbf{v} \in \mathbb{R}^{F}$
according to
\begin{equation}
\mathbf{x}_{i}^{l+1} = \mathbf{u}^l \circ \mathbf{x}_{i}^{l} + \mathbf{W}^l\sigma(\mathbf{v}_{i}^{l}) + \mathbf{b}^l
\label{eq:interaction_block_final}
\end{equation}
where $\mathbf{u}^{l}, \mathbf{b}^{l} \in \mathbb{R}^{F}$ and
$\mathbf{W}^{l} \in \mathbb{R}^{F \times F}$ are parameters and
`$\circ$' denotes the Hadamard (entrywise) product (see
Fig.~\ref{fig:nn_architecture}C). The gating vector $\mathbf{u}$,
inspired by the gated recurrent unit,\cite{cho2014learning} allows
individual entries of the feature vector to be damped or reinforced
during the update.  The final message $\mathbf{v}$ used in
Eq.~\ref{eq:interaction_block_final} is obtained by passing a
proto-message ${\tilde{\mathbf{v}}}$ through $N_{\rm residual}^{\rm
  interaction}$ residual blocks (Eq.~ \ref{eq:residual_block}), where
${\tilde{\mathbf{v}}}$ is given by
\begin{equation}
\tilde{\mathbf{v}}_{i}^{l} = \sigma\left(\mathbf{W_I}^{l}\sigma(\mathbf{x}_{i}^{l}) + \mathbf{b_I}^{l}\right) + \sum_{j\neq i} \mathbf{G}^l\mathbf{g}(r_{ij}) \circ\sigma\left(\mathbf{W_J}^{l}\sigma(\mathbf{x}_{j}^{l}) + \mathbf{b_J}^{l}\right)
\label{eq:interaction_block_intermediate}
\end{equation}
and $\mathbf{W_I}^{l}, \mathbf{W_J}^{l} \in \mathbb{R}^{F \times F}$, $\mathbf{b_I}^{l}, \mathbf{b_J}^{l} \in \mathbb{R}^{F}$ and $\mathbf{G}^{l} \in \mathbb{R}^{F\times K}$ are parameters and $r_{ij}$ denotes the Euclidean distance between atoms $i$ and $j$. The vector $\mathbf{g}(r_{ij}) = \left[g_1(r_{ij})\ \cdots\  g_K(r_{ij}) \right]^\mathsf{T}$ is composed of the values of $K$ radial basis functions of the form
\begin{equation}
g_k(r_{ij}) = \phi(r_{ij})\cdot\exp\left(-\beta_k\left(\exp(-r_{ij}) - \mu_k\right)^2\right)
\label{eq:radial_basis_function}
\end{equation}
where $\mu_k,\beta_k \in \mathbb{R}_{>0}$ are parameters that specify centre and width of $g_k(r_{ij})$, respectively, 
and $\phi(r_{ij})$ is a smooth cut-off function given by\cite{ebert2003texturing}
\begin{equation}
\phi(r_{ij}) = \begin{cases}
1 - 6\left(\dfrac{r_{ij}}{r_{\rm cut}}\right)^5 + 15\left(\dfrac{r_{ij}}{r_{\rm cut}}\right)^4 - 10\left(\dfrac{r_{ij}}{r_{\rm cut}}\right)^3 & r_{ij} < r_{\rm cut}\\
0 & r_{ij} \geq r_{\rm cut}\\
\end{cases}
\label{eq:cut-off_function}
\end{equation}
that ensures continuous behaviour when an atom enters or leaves the
cut-off sphere. The vector $\mathbf{G}^{l}\mathbf{g}(r_{ij}) \in
\mathbb{R}^{F}$ takes the role of a learnable attention
mask\cite{mnih2014recurrent} that selects different features based on
the pairwise distance $r_{ij}$ between atoms. Note that the Gaussian
in Eq.~\ref{eq:radial_basis_function} takes $\exp(-r_{ij})$ instead of
$r_{ij}$ as its argument, which biases attention masks towards a
functional form that decays exponentially with $r_{ij}$ (see
Fig.~\ref{fig:radial_basis_functions}). Such a bias is meaningful for
a chemical system, as it entails the physical knowledge that bound
state wave functions in two-body systems decay exponentially. Since
only pairwise distances are used in
Eq.~\ref{eq:interaction_block_intermediate}, the output of an
interaction block is automatically translationally and rotationally
invariant, while the commutative property of summation ensures
permutational invariance.\cite{o1973exponential}

\paragraph{Output block}
Each output block passes the atomic features through $N^{\rm
  output}_{\rm residual}$ additional residual blocks and computes the
output $\mathbf{y}_{i}^{m} \in \mathbb{R}^{n_{\rm out}}$ of module $m$
for atom $i$ by a linear transformation of the activated features
according to
\begin{equation}
\mathbf{y}_{i}^{m} = \mathbf{W}_{\rm out}^{m}\sigma(\mathbf{x}_{i}^{l}) + \mathbf{b}_{\rm out}^{m}
\label{eq:output_block}
\end{equation}
where $\mathbf{W}_{\rm out}^{m} \in \mathbb{R}^{F \times n_{\rm out}}$
and $\mathbf{b}_{\rm out}^{m} \in \mathbb{R}^{n_{\rm out}}$ are
parameters. How many entries the output vector $\mathbf{y}_{i}^{m}$
has depends on how many atomic properties are predicted at once. In
this work, two variants are considered: The first version predicts
atomic energy contributions along with atomic partial charges
(i.e.\ $\mathbf{y}_{i}^{m} = \left[ E_{i}^{m}
  \ q_{i}^{m}\right]^\mathsf{T}$ and $n_{\rm out} = 2$), whereas the
second version predicts just atomic energy contributions
(i.e.\ $\mathbf{y}_{i}^{m} = \left[ E_{i}^{m} \right]$ and $n_{\rm
  out} = 1$). In principle, other properties could be predicted as
well.

\paragraph{Final prediction}
The final atomic properties $\mathbf{y}_i$ are obtained by summing the
contributions of the individual modules according to
\begin{equation}
\mathbf{y}_{i} = \mathbf{s}_{Z_i}\circ\left(\sum_{m=1}^{N_{\rm module}}\mathbf{y}_{i}^{m}\right) + \mathbf{c}_{Z_i}
\label{eq:final_prediction}
\end{equation}
where $\mathbf{s}_{Z}, \mathbf{c}_{Z} \in \mathbb{R}^{n_{\rm out}}$
are learnable element-specific scale and shift parameters depending on
the nuclear charge $Z_i$ of atom $i$. The scaling and shifting of the
output decouples the values of other parameters from the numeric range
of target properties, which depends mainly on the chosen system of
units. Element-specific (instead of global) parameters are motivated
by a previous observation indicating that atomic properties of
distinct elements can span vastly different
ranges.\cite{unke2018reactive}

The final prediction for the total energy of a system of interest
composed of $N$ atoms is obtained by summation of the atomic energy
contributions $E_i$:
\begin{equation}
E = \sum\limits_{i=1}^{N} E_i
\label{eq:potential_energy_no_longrange}
\end{equation}
A potential shortcoming of Eq.~\ref{eq:potential_energy_no_longrange}
is the fact that all long-range interactions contributing to $E$
beyond the cut-off radius $r_{\rm cut}$ cannot be properly accounted
for. As long as $r_{\rm cut}$ is chosen sufficiently large, this is
not an issue. However, in order to account for electrostatic
interactions which decay with the inverse of the distance, a large
cut-off would be necessary and reduce the computational efficiency of
the model. Fortunately, their functional form is known and they can be
explicitly included when computing $E$. Other types of long-range
interactions for which the functional form is also known analytically,
for example dispersion corrections like
DFT-D3,\cite{grimme2010consistent} can be included as well.

\begin{figure}[htpb]
\centering
\includegraphics[width=0.75\textwidth]{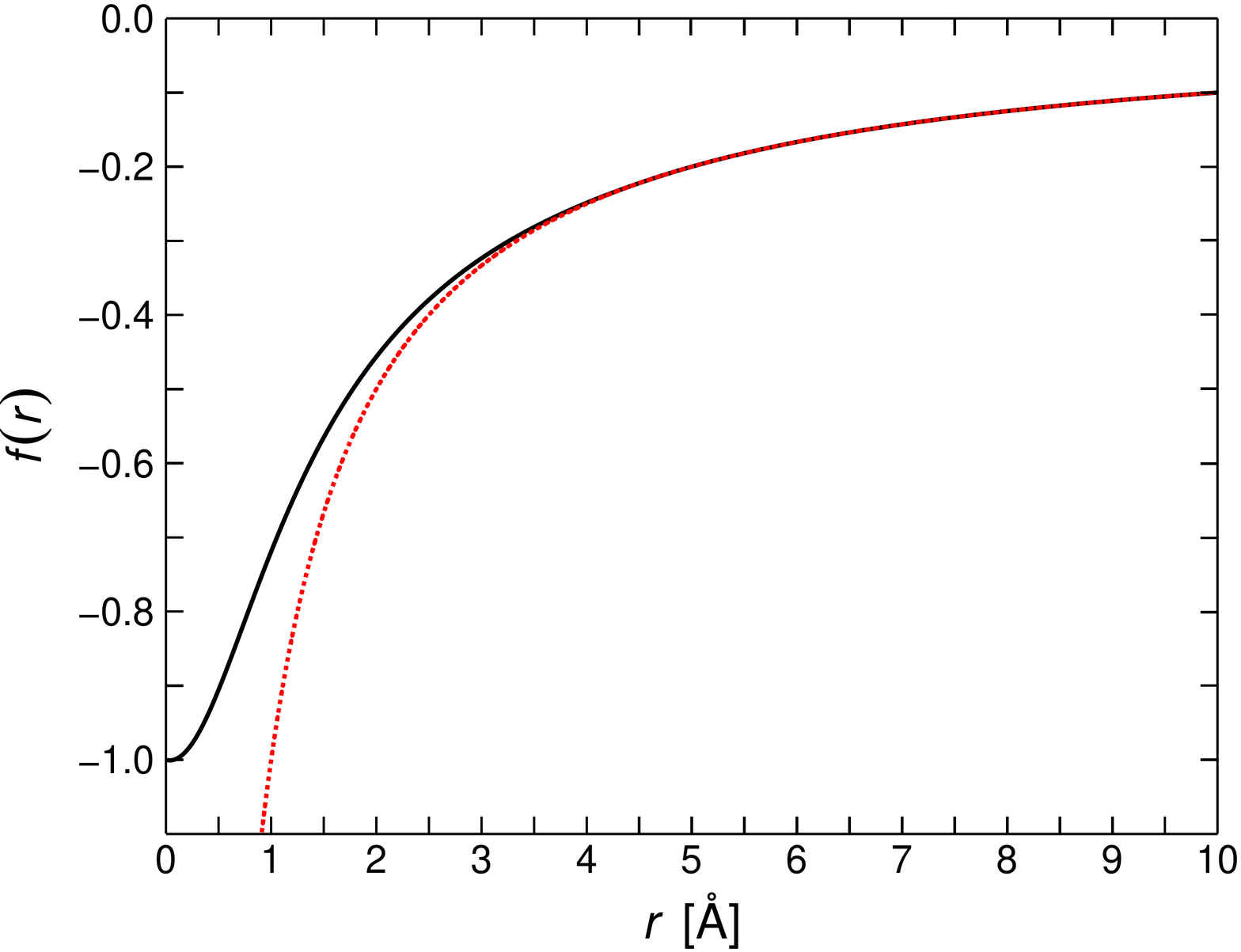}
\caption{Coulomb interaction. The dotted red curve shows the correct
  $\frac{1}{r}$ dependence of Coulomb's law, whereas the solid black
  curve shows $\chi(r)$ (see
  Eq. \ref{eq:shielded_coulomb_interaction}) for $r_{\rm cut} =
  10$\ \AA. Note that for $r > \frac{r_{\rm cut}}{2}$, both curves are
  identical by construction.}
\label{fig:coulomb_interaction}
\end{figure}

Because of the shortcomings of
Eq.~\ref{eq:potential_energy_no_longrange}, for a \nn\ model that also
predicts atomic partial charges, $E$ is calculated by
\begin{equation}
E = \sum\limits_{i=1}^{N} E_i + k_e\sum_{i=1}^{N}\sum_{j>i}^{N}\tilde{q}_{i}\tilde{q}_{j} \chi(r_{ij}) + E_{\rm D3}
\label{eq:potential_energy_with_longrange}
\end{equation}
instead, where $E_{\rm D3}$ is the DFT-D3 dispersion
correction,\cite{grimme2010consistent} $k_{e}$ is Coulomb's constant,
$\tilde{q}_{i}$ and $\tilde{q}_{j}$ are corrected partial charges (see
Eq.~\ref{eq:corrected_partial_charges}) of atoms $i$ and $j$, and
$\chi(r_{ij})$ is given by
\begin{equation}
\chi(r_{ij}) = \phi(2r_{ij})\frac{1}{\sqrt{r_{ij}^2+1}} + \left(1- \phi(2r_{ij})\right)\frac{1}{r_{ij}}
\label{eq:shielded_coulomb_interaction}
\end{equation}
where $\phi(r_{ij})$ is the cut-off function given by
Eq.~\ref{eq:cut-off_function}. Here, $\chi(r_{ij})$ smoothly
interpolates between the correct $r_{ij}^{-1}$ dependence of Coulomb's
law at long-range and a damped term at small distances to avoid the
singularity at $r_{ij}=0$ (see
Fig.~\ref{fig:coulomb_interaction}). The corrected partial charges
$\tilde{q}_i$ are obtained from the partial charges $q_i$ predicted by
the neural network (Eq.~\ref{eq:output_block}) according to
\begin{equation}
\tilde{q}_i = q_i - \frac{1}{N}\left(\sum_{j=1}^{N}q_j - Q\right)
\label{eq:corrected_partial_charges}
\end{equation}
where $Q$ is the total charge of the system. As neural networks are a
purely numerical algorithm, it is not guaranteed \textit{a priori}
that the sum of all predicted atomic partial charges $q_i$ is equal to
the total charge $Q$ (although the result is usually very close when
the neural network is properly trained), so a correction scheme like
Eq.~\ref{eq:corrected_partial_charges} is necessary to guarantee
charge conservation.

It should be pointed out that the summation over all atom pairs when
evaluating the long-range interactions in
Eq.~\ref{eq:potential_energy_with_longrange} makes the evaluation of
$E$ scale quadratically with system size. Fortunately, various schemes
to recover linear scaling are described in the literature, e.g.\ Ewald
summation\cite{darden1993particle} or cut-off
methods\cite{steinbach1994new}, and can be applied without
modification.

Note that the concept of augmenting a neural network potential with
long range interactions is not novel and was first proposed in
[\citenum{artrith2011high}]. However, most previous works use a
separately trained neural network to predict atomic partial
charges,\cite{artrith2011high,yao2018tensormol} in contrast to the
present work, which uses a single neural network to learn both, atomic
energy contributions and partial charges (see
Eqs.~\ref{eq:output_block}~and~\ref{eq:final_prediction}). Aside from
computational advantages (only a single neural network needs to be
trained and evaluated), shared feature representations in such
``multi-task learning'' are believed to increase the generalization
capabilities (transferability) of a
model.\cite{thrun1996learning,caruana1997multitask,baxter2000model}

Apart from allowing the computation of long-range electrostatic
interactions, the partial charges $\tilde{q}_i$ can also be used to
predict the electric dipole moment $\mathbf{p}$ of a structure
according to
\begin{equation}
\mathbf{p} = \sum_{i=1}^{N} \tilde{q}_i\mathbf{r}_i
\end{equation}
where $\mathbf{r}_i$ are the Cartesian coordinates of atom $i$. The
ability to predict $\mathbf{p}$ is useful for example for the
calculation of infrared
spectra.\cite{gastegger2017machine,nebgen2018transferable}

Finally, analytical derivatives of $E$ (see
Eqs.~\ref{eq:potential_energy_no_longrange}~and~\ref{eq:potential_energy_with_longrange})
with respect to the Cartesian coordinates
$\{\mathbf{r}_1,\dots,\mathbf{r}_N\}$ of the atoms, for example in
order to derive the forces $\mathbf{F}_i$ acting on each atom $i$, are
readily obtained by reverse mode automatic
differentiation.\cite{baydin2018automatic}

\paragraph{Hyperparameters}
The \nn\ architecture can be tuned by hyperparameters that control
width and depth of the neural network (see
Fig.~\ref{fig:nn_architecture}). While it would be possible to
optimize hyperparameters for individual learning tasks, for example
via a grid search, it was found that this is not necessary for good
performance. For simplicity, all models used in this work share the
same architecture with the hyperparameters summarized in
Table~\ref{tab:hyperparameters}, unless specified otherwise.
\begin{table}[htbp]
\captionsetup{width=0.85\textwidth}
\caption{Hyperparameters of all models used in this work (unless specified otherwise).}
\label{tab:hyperparameters}
	
\centering
\begin{tabular}{ l l l }
\toprule
\textbf{hyperparameter} & \textbf{value} & \textbf{significance} \\
\midrule
$F$  & \num{128} & dimensionality of feature space \\
\addlinespace[1pt]
$K$  & \num{64} & number of radial basis functions \\
\addlinespace[1pt]
$N_{\rm module}$ & \num{5} & number of stacked modular building blocks\\
\addlinespace[1pt]
$N_{\rm residual}^{\rm atomic}$ & \num{2} & number of residual blocks for atom-wise refinements\\
\addlinespace[1pt]
$N_{\rm residual}^{\rm interaction}$ & \num{3} & number of residual blocks for refinements of proto-message\\
\addlinespace[1pt]
$N_{\rm residual}^{\rm output}$ & \num{1} & number of residual blocks in output blocks\\
\addlinespace[1pt]
$r_{\rm cut}$ & 10 \AA & cut-off radius for interactions in the neural network\\
\bottomrule
\end{tabular}
\end{table}

\subsection{Training}
\label{sec:training}
Before the training of a neural network starts, its parameters need to
be initialized. All entries of the embedding vectors $\mathbf{e}_Z$
are initialized with random values uniformly distributed between
$-\sqrt{3}$ and $\sqrt{3}$ (such that they have unit expected
variance) and the weight matrices $\mathbf{W}$,$\mathbf{W_I}$ and
$\mathbf{W_J}$ are initialized to random orthogonal matrices with
entries scaled such that their variance corresponds to the value
recommended in the Glorot initialization
scheme.\cite{glorot2010understanding} The entries of all bias vectors
$\mathbf{b}$, $\mathbf{b_I}$, $\mathbf{b_J}$, $\mathbf{b}_{\rm out}$
and matrices $\mathbf{G}$ and $\mathbf{W}_{\rm out}$ are initialized
to zero, whereas the entries of the gating vectors $\mathbf{u}$ are
initialized to one. The centres $\mu_k$ of the of the radial basis
functions are set to $K$ equally spaced values between $\exp(-r_{\rm
  cut})$ and $1$ and their widths $\beta_k$ are initialized to
$\left(2K^{-1}\left(1-\exp(-r_{\rm cut})\right)\right)^{-2}$ (see
Fig.~\ref{fig:radial_basis_functions}).

After initialization, the parameters of the neural network are
optimized by minimizing a loss function $\mathcal{L}$ using
AMSGrad\cite{reddi2018convergence} with a learning rate of $10^{-3}$
(other hyperparameters of the optimizer are set to the default values
recommended in [\citenum{reddi2018convergence}]) and a batch size of
\num{32} randomly chosen reference structures. For predicting energies
without long-range augmentation
(Eq.~\ref{eq:potential_energy_no_longrange}) the loss function is
\begin{equation}
\begin{split}
\mathcal{L} = w_{E} \left\lvert E-E^{\rm ref}\right\rvert +
\frac{w_{F}}{3N} \sum_{i=1}^{N}\sum_{\alpha = 1}^{3} \left\lvert
-\frac{\partial E}{\partial r_{i,\alpha}}-F_{i,\alpha}^{\rm ref}
\right\rvert + \mathcal{L}_{\rm nh}
\end{split}
\label{eq:loss_without_charge}
\end{equation}
whereas when energies and charges are predicted
(Eq.~\ref{eq:potential_energy_no_longrange}), the loss function is
\begin{equation}
\begin{split}
\mathcal{L} = w_{E} \left\lvert E-E^{\rm ref}\right\rvert +
\frac{w_{F}}{3N} \sum_{i=1}^{N}\sum_{\alpha = 1}^{3} \left\lvert
-\frac{\partial E}{\partial r_{i,\alpha}}-F_{i,\alpha}^{\rm ref}
\right\rvert + w_{Q}\left\lvert\sum_{i=1}^{N} q_{i} - Q^{\rm
  ref}\right\rvert \\ + \frac{w_{p}}{3}\sum_{\alpha =
  1}^{3}\left\lvert\sum_{i=1}^{N}q_{i} r_{i,\alpha} - p_{\alpha}^{\rm
  ref}\right\rvert + \mathcal{L}_{\rm nh}
\end{split}
\label{eq:loss_with_charge}
\end{equation}

Here, $E^{\rm ref}$ and $Q^{\rm ref}$ are reference energy and total
charge, $p_{\alpha}^{\rm ref}$ are the Cartesian components of the
reference dipole moment $\mathbf{p}^{\rm ref}$, $F_{i,\alpha}^{\rm
  ref}$ are the Cartesian components of the reference force
$\mathbf{F}_i^{\rm ref}$ acting on atom $i$ and $r_{i,\alpha}$ is the
$\alpha$th Cartesian coordinate of atom $i$. The energy prediction $E$
is given either by Eq.~\ref{eq:potential_energy_no_longrange} or
Eq.~\ref{eq:potential_energy_with_longrange}, depending on which
variant of \nn\ is used.

The weighting hyperparameters $w_{E}$, $w_{F}$, $w_{Q}$ and $w_{p}$
determine the relative contribution of the individual error terms to
the loss term. Note that the numeric ranges of the error terms (and
therefore their contribution to $\mathcal{L}$) also depend on the
chosen system of units. For simplicity, weighting hyperparameters are
not optimized for individual learning tasks and instead always set to
$w_{E}=w_{Q}=w_{p}=1$ and $w_{F}=10^2$ (when all quantities are
measured in atomic units). The higher relative weight of force errors
is motivated by the fact that forces alone determine the dynamics of a
chemical system and accurate force predictions are therefore most
important for MD simulations. Note that for datasets where any of the
reference quantities used in
Eqs.~\ref{eq:loss_without_charge}~and~\ref{eq:loss_with_charge} are
not available, the corresponding weight is set to zero.

The term $\mathcal{L}_{\rm nh}$ is a ``non-hierarchicality penalty'',
inspired by a similar regularization method introduced in
[\citenum{lubbers2018hierarchical}], given either by
\begin{equation}
\mathcal{L}_{\rm nh} = \frac{\lambda_{\rm
    nh}}{N}\sum_{i=1}^{N}\sum_{m=2}^{N_{\rm module}}
\frac{(E_i^m)^2}{(E_i^m)^2+(E_i^{m-1})^2}
\label{eq:nhloss_without_charge}
\end{equation}
or
\begin{equation}
\mathcal{L}_{\rm nh} = \frac{\lambda_{\rm
    nh}}{N}\sum_{i=1}^{N}\sum_{m=2}^{N_{\rm module}}
\frac{1}{2}\left[\frac{(E_i^m)^2}{(E_i^m)^2+(E_i^{m-1})^2} +
  \frac{(q_i^m)^2}{(q_i^m)^2+(q_i^{m-1})^2}\right]
\label{eq:nhloss_with_charge}
\end{equation}
depending on which variant of \nn\ is used, and $\lambda_{\rm nh}$ is
the corresponding regularization hyperparameter. The $\mathcal{L}_{\rm
  nh}$ term penalizes when the predictions of individual modules do
not decay with increasing depth in the hierarchy. Since deeper feature
representations of atoms capture increasingly higher-order
interactions, such a regularization is motivated by the fact that
higher-order terms in many-body expansions of the energy are known to
decay rapidly in magnitude. For simplicity, $\lambda_{\rm nh}$ is not
tuned for individual learning tasks and instead always set to
$10^{-2}$.

During training, an exponential moving average of all parameter values
is kept using a decay rate of $0.999$. Overfitting is prevented using
early stopping:\cite{prechelt2012early} After every epoch (one pass
over all reference structures in the training set), the loss function
is evaluated on a validation set of reference structures using the
parameters obtained from the exponential moving average. After
training, the model that performed best on the validation set is
selected. Since the validation set is used indirectly during the
training procedure, the performance of the final models (see
section~\ref{sec:results}) is always measured on a separate test set.

Note that only true quantum mechanical observables, such as total
energy, forces or dipole moments, are used as reference when training
the neural network (see Eqs.~\ref{eq:loss_without_charge}
and~\ref{eq:loss_with_charge}). While it would also be possible to
train directly on atomic energies and partial charges obtained using a
decomposition
method,\cite{hirshfeld1977bonded,blanco2005interacting,francisco2006molecular,mitoraj2009combined}
such schemes are essentially arbitrary and it is unclear whether the
corresponding decompositions are meaningful. Further, it is not always
guaranteed that the quantities obtained from such methods vary
smoothly when the molecular geometry changes, which makes it difficult
for a neural network to learn them. By only relying on quantum
mechanical observables, the model automatically learns to perform a
smooth decomposition in a data-driven way.

\subsection{Dataset generation}
\label{sec:dataset_generation}
In the following, the generation of two new benchmark datasets is
described, which probe chemical reactivity, long-range electrostatics,
and many-body intermolecular interactions.

\paragraph{S$_{\rm N}$2 reactions}
\label{sec:dataset_generation:sn2}
The S$_{\rm N}$2 reactions dataset probes chemical reactions of the
kind X$^{-}$ + H$_3$C--Y $\rightarrow$ X--CH$_3$ + Y$^{-}$ and
contains structures for all possible combinations of X,Y $\in$ \{F,
Cl, Br, I\}. The reactions of methyl halides with halide anions are a
prototypical examples for chemical reactions and have been studied
extensively both
experimentally\cite{tanaka1976gas,o1994measurements,li1996high,deturi1997translational,angel2003gas}
and
theoretically.\cite{deng1994potential,hu1995deuterium,glukhovtsev1996gas,lee2002high,liu2010steric,stei2016influence}
It consists of different geometries for the high-energy transition
regions, ion-dipole bound state complexes and long-range ($>10$~\AA)
interactions of CH$_3$X molecules with Y$^{-}$ ions. The dataset also
includes various structures for several smaller molecules that can be
formed in fragmentation reactions, such as CH$_3$X, HX, CHX or
CH$_2$X$^{-}$ with X $\in$ \{F, Cl, Br, I\}, as well as geometries for
H$_2$, CH$_2$, CH$_3^+$ and XY interhalogen compounds for all possible
combinations of X,Y $\in$ \{F, Cl, Br, I\}. In total, the dataset
provides reference energies, forces, and dipole moments for
\num{452709} structures calculated at the DSD-BLYP-D3(BJ)/def2-TZVP
level of
theory.\cite{kozuch2010dsd,grimme2010consistent,grimme2011effect,weigend2005balanced}

Different conformations for each species present in the S$_{\rm N}$2
reactions dataset were sampled by running MD simulations at a
temperature of \num{5000}~K with a time step of $0.1$~fs using the
Atomic Simulation Environment (ASE).\cite{larsen2017atomic} The
necessary forces were obtained with the semi-empirical PM7
method\cite{stewart2013optimization} implemented in
MOPAC2016.\cite{mopac2016} Structures were saved every \num{10} steps
and for each of them, reference energies, forces and dipole moments
were calculated at the
DSD-BLYP-D3(BJ)/def2-TZVP\cite{kozuch2010dsd,grimme2010consistent,grimme2011effect,weigend2005balanced}
level of theory using the ORCA 4.0.1
code.\cite{neese2012orca,neese2018software} The DSD-BLYP functional is
one of the best performing double hybrid methods in the GMTKN55
benchmark.\cite{goerigk2017look} All MD simulations were started from
the PM7-optimized geometries. For the reaction complexes
$[\mathrm{XCH}_3\mathrm{Y}]^-$, MD simulations were randomly started
either from the respective van der Waals complexes, or, in order to
better sample the transition regions, from the transition state
(calculated with PM7) of the respective reaction. Further, long-range
interactions were sampled by choosing a random conformation from the
CH$_3$X MD simulations and randomly placing an ion Y$^-$ in the
vicinity of the CH$_3$X molecule such that its distance to any other
atom is at most $16$~\AA. Table~\ref{tab:sn2_dataset_composition}
lists the number of conformations for each species.

\begin{table}[htbp]
\footnotesize
\centering
\caption[Number of structures in the S$_{\rm N}$2 reactions
  dataset]{Number of structures for each species present in the
  S$_{\rm N}$2 reactions dataset.}
\label{tab:sn2_dataset_composition}
\begin{tabular}{l c }
	\toprule
		\textbf{species} & \textbf{count} \\
	\midrule
	$[\mathrm{FCH}_3\mathrm{Cl}]^-$ \phantom{$]_{32}^{+-}$} & \num{44501}\\
	\addlinespace[1pt]
	$[\mathrm{FCH}_3\mathrm{Br}]^-$ \phantom{$]_{32}^{+-}$} & \num{44501}\\
	\addlinespace[1pt]
	$[\mathrm{FCH}_3\mathrm{I}]^-$ \phantom{$]_{32}^{+-}$} & \num{44501}\\
	\addlinespace[1pt]
	$[\mathrm{ClCH}_3\mathrm{Br}]^-$ \phantom{$]_{32}^{+-}$} & \num{44501}\\
	\addlinespace[1pt]
	$[\mathrm{ClCH}_3\mathrm{I}]^-$ \phantom{$]_{32}^{+-}$} & \num{44501}\\
	\addlinespace[1pt]
	$[\mathrm{BrCH}_3\mathrm{I}]^-$ \phantom{$]_{32}^{+-}$}  & \num{44501}\\
	\addlinespace[1pt]
	$[\mathrm{FCH}_3\mathrm{F}]^-$ \phantom{$]_{32}^{+-}$} & \num{24801}\\
	\addlinespace[1pt]
	$[\mathrm{ClCH}_3\mathrm{Cl}]^-$ \phantom{$]_{32}^{+-}$} & \num{24801}\\
	\addlinespace[1pt]
	$[\mathrm{BrCH}_3\mathrm{Br}]^-$ \phantom{$]_{32}^{+-}$} & \num{24801}\\
	\addlinespace[1pt]
	$[\mathrm{ICH}_3\mathrm{I}]^-$ \phantom{$]_{32}^{+-}$} & \num{24801}\\
	\addlinespace[1pt]
	$\mathrm{CH}_3\mathrm{F}$ \phantom{$]_{32}^{+-}$} & \num{3500}\\
	\addlinespace[1pt]
	$\mathrm{CH}_3\mathrm{Cl}$ \phantom{$]_{32}^{+-}$} & \num{3500}\\
	\addlinespace[1pt]
	$\mathrm{CH}_3\mathrm{Br}$ \phantom{$]_{32}^{+-}$} & \num{3500}\\
	\bottomrule
\end{tabular}
\begin{tabular}{l c }
	\toprule
	\textbf{species} & \textbf{count} \\
	\midrule
	$\mathrm{CH}_3\mathrm{I}$ \phantom{$]_{32}^{+-}$} & \num{3500}\\
	\addlinespace[1pt]
	$\mathrm{CH}_3^+$ \phantom{$]_{32}^{+-}$}  & \num{3500}\\
	\addlinespace[1pt]
	$\mathrm{CH}_2\mathrm{F}^-$ \phantom{$]_{32}^{+-}$} & \num{3500}\\
	\addlinespace[1pt]
	$\mathrm{CH}_2\mathrm{Cl}^-$ \phantom{$]_{32}^{+-}$} & \num{3500}\\
	\addlinespace[1pt]
	$\mathrm{CH}_2\mathrm{Br}^-$ \phantom{$]_{32}^{+-}$}  & \num{3500}\\
	\addlinespace[1pt]
	$\mathrm{CH}_2\mathrm{I}^-$ \phantom{$]_{32}^{+-}$} & \num{3500}\\
	\addlinespace[1pt]
	$\mathrm{CH}_2$ \phantom{$]_{32}^{+-}$} & \num{3500}\\
	\addlinespace[1pt]
	$\mathrm{CHF}$ \phantom{$]_{32}^{+-}$} & \num{3500}\\
	\addlinespace[1pt]
	$\mathrm{CHCl}$ \phantom{$]_{32}^{+-}$} & \num{3500}\\
	\addlinespace[1pt]
	$\mathrm{CHBr}$ \phantom{$]_{32}^{+-}$}  & \num{3500}\\
	\addlinespace[1pt]
	$\mathrm{CHI}$ \phantom{$]_{32}^{+-}$} & \num{3500}\\
	\addlinespace[1pt]
	$\mathrm{H}_2$ \phantom{$]_{32}^{+-}$} & \num{3500}\\
	\addlinespace[1pt]
	$\mathrm{HF}$ \phantom{$]_{32}^{+-}$} & \num{3500}\\
	\bottomrule
\end{tabular}
\begin{tabular}{l c }
	\toprule
	\textbf{species} & \textbf{count} \\
	\midrule
	$\mathrm{HCl}$ \phantom{$]_{32}^{+-}$} & \num{3500}\\
	\addlinespace[1pt]
	$\mathrm{HBr}$ \phantom{$]_{32}^{+-}$} & \num{3500}\\
	\addlinespace[1pt]
	$\mathrm{HI}$ \phantom{$]_{32}^{+-}$} & \num{3500}\\
	\addlinespace[1pt]
	$\mathrm{F}_2$ \phantom{$]_{32}^{+-}$} & \num{2000}\\
	\addlinespace[1pt]
	$\mathrm{FCl}$ \phantom{$]_{32}^{+-}$} & \num{2000}\\
	\addlinespace[1pt]
	$\mathrm{FBr}$ \phantom{$]_{32}^{+-}$} & \num{2000}\\
	\addlinespace[1pt]
	$\mathrm{FI}$ \phantom{$]_{32}^{+-}$} & \num{2000}\\
	\addlinespace[1pt]
	$\mathrm{Cl}_2$ \phantom{$]_{32}^{+-}$} & \num{1999}\\
	\addlinespace[1pt]
	$\mathrm{ClBr}$ \phantom{$]_{32}^{+-}$} & \num{2000}\\
	\addlinespace[1pt]
	$\mathrm{ClI}$ \phantom{$]_{32}^{+-}$} & \num{2000}\\
	\addlinespace[1pt]
	$\mathrm{Br}_2$ \phantom{$]_{32}^{+-}$} & \num{2000}\\
	\addlinespace[1pt]
	$\mathrm{BrI}$ \phantom{$]_{32}^{+-}$} & \num{2000}\\
	\addlinespace[1pt]
	$\mathrm{I}_2$ \phantom{$]_{32}^{+-}$} & \num{2000}\\
	\bottomrule
\end{tabular}
\end{table}

\paragraph{Solvated protein fragments}
\label{sec:dataset_generation:solvated_protein_fragments}
The solvated protein fragments dataset probes many-body intermolecular
interactions between ``protein fragments'' and water molecules, which
are important for the description of many biologically relevant
condensed phase systems. It contains structures for all possible
``amons''\cite{huang2017chemical} (hydrogen-saturated covalently
bonded fragments) of up to eight heavy atoms (C, N, O, S) that can be
derived from chemical graphs of proteins containing the \num{20}
natural amino acids connected via peptide bonds or disulfide
bridges. For amino acids that can occur in different charge states due
to (de-)protonation (i.e.\ carboxylic acids that can be negatively
charged or amines that can be positively charged), all possible
structures with up to a total charge of $\pm2e$ are included. These
structures are augmented with solvated variants containing a varying
number of water molecules such that the total number of heavy atoms
does not exceed \num{21}. The dataset also contains randomly sampled
dimer interactions of protein fragments, as well as structures of pure
water with up to \num{40} molecules. For all included structures,
several conformations sampled with MD simulations at \num{1000}~K are
included. In total, the dataset contains reference energies, forces
and dipole moments for \num{2731180} structures calculated at the
revPBE-D3(BJ)/def2-TZVP level of
theory.\cite{zhang1998comment,grimme2010consistent,grimme2011effect,weigend2005balanced}
On average, the structures contain \num{21} atoms (with a maximum of
\num{120} atoms) and consist of 63\% hydrogen, 19\% carbon, 12\%
oxygen, 5\% nitrogen, and 1\% sulfur atoms.

\noindent
The different structures in the dataset were constructed as follows:
All amons with up to eight heavy atoms (C, N, O, S) were constructed
according to the method described in [\citenum{huang2017chemical}] for
all possible chemical graphs of proteins containing the \num{20}
natural amino acids connected via peptide bonds or disulfide
bridges. For amons derived from amino acids that can occur in
different charge states due to (de-)protonation, variants for all
charge states are included. This results in \num{2307} different
molecules. In order to sample interactions with solvent molecules, the
amon structures were augmented by randomly placing up to \num{20}
water molecules in their vicinity, such that the total number of heavy
atoms does not exceed \num{21}. This results in \num{29991} additional
structures. Further, other intermolecular interactions were sampled by
generating all possible dimers from amons with up to \num{3} heavy
atoms, resulting in \num{867} possible combinations. Important
interactions between different amino acids were included by adding
sidechain-sidechain and backbone-backbone complexes from the
BioFragment Database\cite{burns2017biofragment} (\num{3480}
structures). Further, interactions in pure water were sampled by
constructing water clusters with up to \num{21} molecules. Each water
cluster is complemented by a variant with an additional proton, as
well as with a variant lacking one proton in order to sample the
different possible charge states of water. This results in \num{24}
additional structures.

All structures were optimized using the semi-empirical PM7
method\cite{stewart2013optimization} implemented in
MOPAC2016.\cite{mopac2016} Starting from the optimized geometry,
\num{100} different conformations for each structure were sampled by
running MD simulations (at the same level of theory) at a temperature
of \num{1000}~K with a time step of $0.1$~fs using the
ASE.\cite{larsen2017atomic} Structures were saved every \num{10} steps
and for each of them, reference energies, forces and dipole moments
were calculated at the
revPBE-D3(BJ)/def2-TZVP\cite{zhang1998comment,grimme2010consistent,grimme2011effect,weigend2005balanced}
level of theory using the ORCA 4.0.1
code.\cite{neese2012orca,neese2018software} The revPBE functional is
one of the best performing GGA functionals in the GMTKN55
benchmark.\cite{goerigk2017look}

While this initial data already covers many different chemical
situations, it is not guaranteed that the contained structures cover
chemical and configurational space sufficiently well to account for
all situations that might be relevant in MD simulations. For this
reasons, the initial dataset was iteratively augmented using an
adaptive sampling
method:\cite{behler2014representing,behler2015constructing} An
ensemble\cite{breiman1996bagging} of three \nn\ models trained (see
section~\ref{sec:training}) on the initial dataset is used to run MD
simulations and all structures for which their predictions deviate by
more than a threshold value (here $1$~kcal~mol$^{-1}$) are
saved. Discrepancies in the predictions are a strong indicator that
the dataset used for training the models does not contain sufficient
information for a particular
conformation.\cite{behler2014representing,behler2015constructing} For
each structure saved in this process, energies, forces and dipole
moments were calculated with the reference \textit{ab initio} method
and added to the dataset. Afterwards, the models were retrained and
the sampling process repeated. In total, the dataset was adaptively
augmented in this way for four times, after which significant
deviations between the predictions were found to be rare. Finally,
energies, forces and dipole moments were calculated with the reference
method for \num{10000} structures of \num{40} water molecules in a
spherical arrangement (obtained by running MD simulations with PM7,
see above) to include training examples similar to bulk phase
water. The final dataset contains data for \num{2731180} structures.

\section{Results}
\label{sec:results}
In this section, \nn\ is applied to various quantum-chemical datasets
that all probe different aspects of chemical space (i.e.\ chemical
and/or conformational degrees of freedom). Apart from the
well-established benchmarks QM9,\cite{ramakrishnan2014quantum}
MD17,\cite{chmiela2017machine} and ISO17,\cite{schutt2017schnet} the
model is also applied to the two new datasets introduced in section
\ref{sec:dataset_generation}.

\paragraph{QM9}
The QM9 dataset\cite{ramakrishnan2014quantum} is a widely used
benchmark for the prediction of several properties of molecules in
equilibrium. It consists of geometric, energetic, electronic, and
thermodynamic properties for $\approx$134k small organic molecules
made up of H, C, O, N, and F atoms. These molecules correspond to the
subset of all species with up to nine heavy atoms (C, O, N, and F) out
of the GDB-17 chemical universe
database.\cite{ruddigkeit2012enumeration} All properties are
calculated at the B3LYP/6-31G(2df,p) level of theory. About 3k
molecules within QM9 fail a geometric consistency check or are
difficult to converge\cite{ramakrishnan2014quantum} and are commonly
removed from the
dataset.\cite{faber2017prediction,gilmer2017neural,schutt2017schnet,lubbers2018hierarchical}
Since the QM9 contains only equilibrium geometries, the benchmark
probes just chemical (but no conformational) degrees of freedom.

Table~\ref{tab:qm9_results} compares the performance for predicting
the energy on the pruned QM9 dataset ($\approx$131k structures) of
various models published in the literature with \nn. Results for
\nn\ are averaged over five independent runs using the same training
set. The performance of \nn\ can be further improved by
\textit{bagging}:\cite{breiman1996bagging} While a single \nn\ model
already improves upon the state-of-the-art, an ensemble of five
\nn\ models (\nn-ens5) brings the error down even further.

\begin{table}[htbp]
\captionsetup{width=0.85\textwidth}
\caption{Mean absolute errors in kcal mol$^{-1}$ for energy
  predictions on the QM9 dataset for several models reported in the
  literature and different training set sizes. Results for \nn\ are
  averaged over five independent runs, whereas \nn-ens5 is the
  performance of an ensemble\cite{breiman1996bagging} of five
  \nn\ models. Best results are shown in bold.}
\label{tab:qm9_results}
	
	\centering
	\begin{tabular}{c c c c c c c}
	\toprule
	$N_{\rm train}$ +  $N_{\rm valid}$ &
	\textbf{enn-s2s-ens5}\cite{gilmer2017neural} &
	\textbf{DTNN}\cite{schutt2017quantum} & \textbf{SchNet}\cite{schutt2017schnet} &  
	\textbf{HIP-NN}\cite{lubbers2018hierarchical} & 
 	\textbf{\nn} & 
 	\textbf{\nn-ens5}\\
	\midrule
	\num{110426} &  0.33 & --&  0.31 & 0.26  &0.19  & \bf 0.14\\
	\addlinespace[1pt]
	\num{100000}    & --  & 0.84  & 0.34 & 0.26  & 0.19 &  \bf 0.14\\
	\addlinespace[1pt]
	\num{50000}   & --  & 0.94 &  0.49 & 0.35  & 0.30 & \bf 0.24\\
	\bottomrule
	\end{tabular}
\end{table}

\paragraph{MD17}
The MD17 dataset\cite{chmiela2017machine} is a collection of
structures, energies and forces from \textit{ab initio} MD simulations
of eight small organic molecules. All trajectories are calculated at a
temperature of \num{500}~K and a resolution of $0.5$~fs using the
PBE+vdW-TS\cite{perdew1996generalized, tkatchenko2009accurate}
electronic structure method. The datasets range in size from 150k to
almost 1M conformations and cover energy ranges between $20$ to
$48$~kcal~mol$^{-1}$ and force components between $266$ to
$570$~kcal~mol$^{-1}$~\AA$^{-1}$. The task is to predict energies (and
forces) using a separate model for each molecule. Since each task is
limited to a single molecule, the MD17 benchmark probes only
conformational (an no chemical) degrees of freedom.

Table~\ref{tab:md17_results} compares the performance of various ML
models published in the literature on the MD17 benchmark with
\nn. Results for \nn\ are averaged over five independent runs using
the same training set and their ensemble
prediction\cite{breiman1996bagging} is also reported (\nn-ens5). When
comparing different models, it should be noted that they use different
subsets of the available data for training:
DTNN\cite{schutt2017quantum} and HIP-NN\cite{lubbers2018hierarchical}
are trained on energies only, GDML\cite{chmiela2017machine} is trained
on forces only, and SchNet and \nn\ are trained on energies and
forces.\cite{schutt2017schnet} \nn-ens5 matches or even improves upon
state-of-the-art performance for all molecules in at least one
category (energy or forces).

\begin{table}[htbp]
\captionsetup{width=0.9\textwidth}
\caption{Mean absolute errors for predictions of energy (in
  kcal~mol$^{-1}$) and forces (in kcal~mol$^{-1}$~\AA$^{-1}$) for
  molecules in the MD17 dataset for several models reported in the
  literature. All models utilize a combined 50k structures for
  training and validation, but using different reference data: DTNN
  and HIP-NN are trained on energies
  only,\cite{schutt2017quantum,lubbers2018hierarchical} GDML is
  trained on forces only,\cite{chmiela2018towards} and SchNet and
  \nn\ are trained on energies and forces.\cite{schutt2017schnet}
  Results for \nn\ are averaged over five independent runs, whereas
  \nn-ens5 is the performance of an ensemble\cite{breiman1996bagging}
  of five \nn\ models. The best results in each category are shown in
  bold.}
\label{tab:md17_results}
	
\centering
\begin{tabular}{c c c c c c c c}
\toprule
	& & 
\textbf{DTNN}\cite{schutt2017quantum} &
\textbf{HIP-NN}\cite{lubbers2018hierarchical}  &
\textbf{GDML}\cite{chmiela2017machine} &
\textbf{SchNet}\cite{schutt2017schnet} &
\textbf{\nn} &
\textbf{\nn-ens5}\\
\midrule

\multirow{ 2}{*}{\bf Aspirin} & \textit{energy} &  --- & --- & 0.13  & \bf 0.12 & \bf 0.12 & \bf 0.12\\
\addlinespace[1pt]
& \textit{forces} & --- & --- & \bf 0.02 & 0.33 & 0.06 & 0.04\\
\midrule

\multirow{ 2}{*}{\bf Benzene} &\textit{energy} & \bf 0.04 & 0.06 & 0.07   & 0.07 &  0.07 & 0.07 \\
\addlinespace[1pt]
& \textit{forces} & --- & --- & 0.24 & 0.17 & 0.15 & \bf 0.14 \\
\midrule

\multirow{ 2}{*}{\bf Ethanol} &\textit{energy} & --- & --- & \bf 0.05 & \bf 0.05 & \bf 0.05 & \bf 0.05\\
\addlinespace[1pt]
& \textit{forces} & --- & --- & 0.09 &0.05 & 0.03 & \bf 0.02\\
\midrule

\multirow{ 2}{*}{\bf Malonaldehyde} & \textit{energy} & 0.19 & 0.09 & 0.08 & 0.08 & \bf 0.07 & \bf 0.07\\
\addlinespace[1pt]
& \textit{forces} & --- & --- & 0.09 & 0.08 & 0.04 & \bf 0.03\\
\midrule

\multirow{ 2}{*}{\bf Naphthalene} & \textit{energy} & --- & --- & 0.12 & \bf 0.11 & 0.12 & 0.12\\
\addlinespace[1pt]
& \textit{forces} & --- & --- & \bf 0.03 & 0.11 & 0.04 & \bf 0.03\\
\midrule

\multirow{ 2}{*}{\bf Salicylic acid} &\textit{energy} & 0.41 & 0.20 & 0.11 & \bf 0.10 & 0.11 & 0.11\\
\addlinespace[1pt]
& \textit{forces} & --- & --- & \bf 0.03 & 0.19 & 0.04 & \bf 0.03\\
\midrule

\multirow{ 2}{*}{\bf Toluene} & \textit{energy} & 0.18 & 0.14 & \bf 0.09 & \bf 0.09 & 0.10 & 0.10\\
\addlinespace[1pt]
& \textit{forces} & --- & --- & 0.05 & 0.09 & \bf 0.03 & \bf 0.03\\
\midrule

\multirow{ 2}{*}{\bf Uracil} & \textit{energy} & --- & --- & 0.11 & \bf 0.10 & \bf 0.10 & \bf 0.10\\
\addlinespace[1pt]
& \textit{forces} & --- & --- & \bf 0.03 & 0.11 & \bf 0.03 & \bf 0.03\\
\midrule

\multicolumn{2}{c}{\textbf{total count of best in class}}  & 1 & 0 & 6 & 6 & 6 & 11\\

\bottomrule
\end{tabular}
\end{table}

\paragraph{ISO17}
The ISO17 dataset\cite{schutt2017schnet} consists of short MD
trajectories of \num{127} isomeric molecules with the composition
C$_{7}$O$_{2}$H$_{10}$ drawn randomly from the largest set of isomers
in QM9. Each trajectory samples 5k conformations at a resolution of
$1$~fs. In total, the dataset contains 635k structures, for which
energies and forces, calculated at the
PBE+vdW-TS\cite{perdew1996generalized, tkatchenko2009accurate} level
of theory, are reported. The task is to predict energies (and forces)
for two different scenarios: In the first variant (known molecules /
unknown conformations), the training set contains $\approx$80\% of all
molecules and conformations (400k structures for training 4k
structures for validation) and the task is to predict the remaining
$\approx$20\% of conformations for the same subset of molecules
present in the training set (101k structures). Thus, the first variant
tests the generalization capabilities of the model for unknown
conformations of previously seen molecules. In the second, more
challenging variant (unknown molecules / unknown conformations), the
training set remains the same, but the task is to predict all 5k
conformations of the $\approx$20\% of molecules not present in the
training set (130k structures). Here, generalization capabilities of
the model are tested for unknown conformations of unknown
molecules. Both variants of the ISO17 benchmark probe chemical and
conformational degrees of freedom.

Table~\ref{tab:iso17_results} compares the performance of
SchNet\cite{schutt2017schnet} to the average performance of five
\nn\ models, as well as their ensemble
prediction\cite{breiman1996bagging} (\nn-ens5) for the two variants of
the ISO17 benchmark. While for the first variant of the benchmark
(known molecules / unknown conformations), \nn\ achieves
state-of-the-art performance on both energies and forces, for the
second variant (unknown molecules / unknown conformations),
\nn\ improves upon SchNet only for force predictions, but performs
slightly worse for energies. This is likely due to the higher relative
weight of force data during the training process (see
section~\ref{sec:training}) and it is possible that the results on
energy could be improved by tuning the corresponding weighting
hyperparameters.

\begin{table}[htbp]
\caption{Mean absolute errors for predictions of energy (in
  kcal~mol$^{-1}$) and forces (in kcal~mol$^{-1}$~\AA$^{-1}$) for the
  two variants of the ISO17 benchmark. \nn\ is compared with the
  performance of SchNet.\cite{schutt2017schnet} Results for \nn\ are
  averaged over five independent runs, whereas \nn-ens5 is the
  performance of an ensemble\cite{breiman1996bagging} of five
  \nn\ models. Best results are shown in bold.}
\label{tab:iso17_results}
\centering
	\begin{tabular}{c c c c c}
	\toprule
	& & 
	\textbf{SchNet}\cite{schutt2017schnet} &
	\textbf{\nn} &
	\textbf{\nn-ens5}\\
	\midrule
	
	{\bf known molecules /} & \textit{energy}& 0.36 & \bf 0.10 & \bf 0.10 \\
	\addlinespace[1pt]
	{\bf unknown conformations  } & \textit{forces} & 1.00 & 0.12 & \bf 0.08 \\
	\midrule
	
	{\bf unknown molecules /} & \textit{energy} & \bf 2.40 & 2.94 & 2.86\\
	\addlinespace[1pt]
	{\bf unknown conformations} 	& \textit{forces} & 2.18 & 1.38 & \bf 1.13 \\
	\bottomrule
\end{tabular}
\end{table}

\paragraph{S$_{\rm N}$2 reactions}
This benchmark dataset was newly generated for the present work, for a
detailed description of the dataset, see
section~\ref{sec:dataset_generation:sn2}. The task is to predict
energies, forces and dipole moments using a single model for all
structures contained in the dataset, testing the generalization
capabilities of the model across chemical and conformational degrees
of freedom, chemical reactions, and challenging long-range
intermolecular interactions.

\noindent
Table~\ref{tab:sn2_results} lists the performance of \nn\ with and
without explicit long-range electrostatic interactions (see
Eqs.~\ref{eq:potential_energy_with_longrange}~and~\ref{eq:potential_energy_no_longrange}). All
models were trained on 400k structures with 5k structures used for
validation. The results in each case are averaged over five
independent runs and the performance of
ensembles\cite{breiman1996bagging} of five \nn\ models is also
reported. Because of the partial charge correction scheme (see
Eq.~\ref{eq:corrected_partial_charges}), the total charge is always
predicted exactly. However, for completeness, the error for the
prediction of total charge (in $e$) using the uncorrected partial
charges is also given.

\begin{table}[htbp]
\captionsetup{width=0.80\textwidth}
\caption{Mean absolute errors for predictions of energy (in
  kcal~mol$^{-1}$), forces (in kcal~mol$^{-1}$~\AA$^{-1}$) and dipole
  moments (in D) for the S$_{\rm N}$2 reactions dataset for \nn\ with
  and without long-range augmentation (see
  Eqs.~\ref{eq:potential_energy_no_longrange}~and~
  \ref{eq:potential_energy_with_longrange}). Results for \nn\ are
  averaged over five independent runs, whereas \nn-ens5 is the
  performance of an ensemble\cite{breiman1996bagging} of five
  models. The best results in each category are shown in bold. Note
  that because of the partial charge correction scheme (see
  Eq.\ref{eq:corrected_partial_charges}), total charge is always
  predicted exactly. However, for completeness, the error for the
  prediction of total charge (in $e$) using the uncorrected partial
  charges is also reported.}
	\label{tab:sn2_results}
	\centering
	\begin{tabular}{c c c S S }
		\toprule
		& 
		\textbf{\nn} &
		\textbf{\nn-ens5} &
		\multicolumn{1}{c}{\textbf{\nn}} &
		\multicolumn{1}{c}{\textbf{\nn-ens5}}  \\
		&
		\textbf{without long-range}&
		\textbf{without long-range}&
		\multicolumn{1}{c}{\textbf{with long-range}} &
		\multicolumn{1}{c}{\textbf{with long-range}}\\
		\midrule
		\textit{energy} & 0.071 & 0.070 & 
		 \boldentry{3.3}{0.009}& \boldentry{2.3}{0.009} \\
		\addlinespace[1pt]
		\textit{forces} & 0.035 & 0.032 & 0.012 & \boldentry{2.3}{0.009} \\
		\addlinespace[1pt]
		\textit{dipole} & --- & --- & 0.0044 & \boldentry{3.4}{0.0042}  \\
		\addlinespace[1pt]
		\textit{charge} & --- & --- & 0.00023 & \boldentry{4.5}{0.00019}  \\
		\bottomrule
	\end{tabular}
\end{table}

\begin{figure}[htbp]
\centering
\captionsetup{width=1.0\textwidth}
\includegraphics[width=0.8\textwidth]{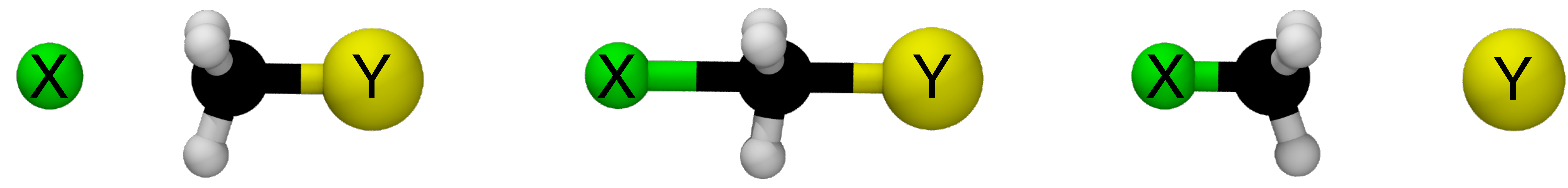}
\includegraphics[width=\textwidth]{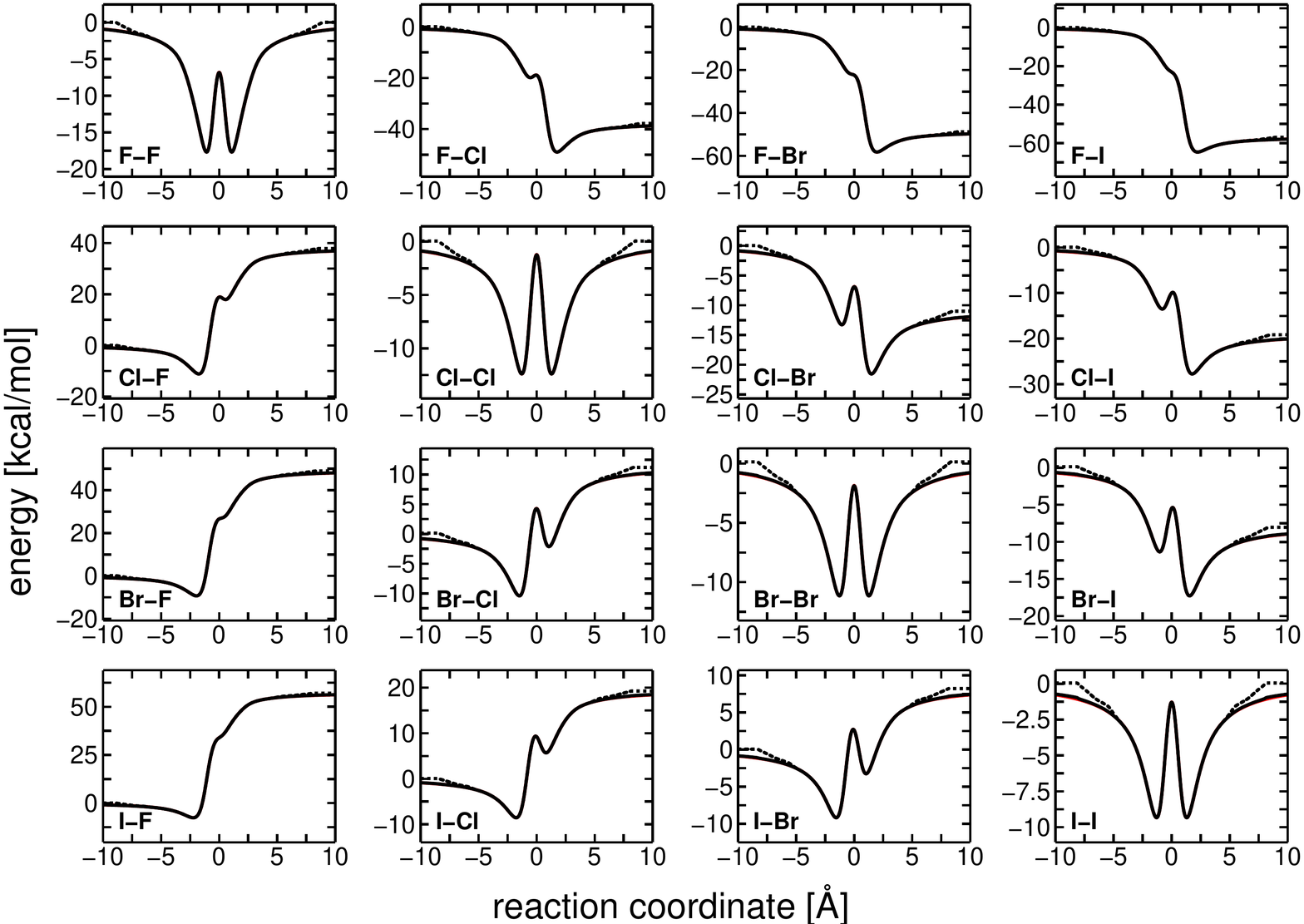}
\caption{Minimum energy paths (MEPs) for S$_{\rm N}$2 reactions of the
  kind X$^{-}$ + H$_3$C-Y $\rightarrow$ X-CH$_3$ + Y$^{-}$ along the
  reaction coordinate defined by the distance difference $r_{\rm
    CY}-r_{\rm CX}$, calculated using the \nn-ens5 model with (solid
  black line) and without (dotted black line) explicit long-range
  interactions. The solid red line (mostly occluded by the solid black
  line) depicts the MEP calculated using the reference method. Each
  panel shows a different combination X-Y with X, Y $\in$ \{F, Cl, Br,
  I\} ($y$-axes are scaled individually for each combination to
  increase visibility). The model including long-range interactions in
  its functional form (Eq.~\ref{eq:potential_energy_with_longrange})
  is virtually identical to the reference method for all values of the
  reaction coordinate (apart from small deviations in the
  asymptotics), whereas the model without long-range interactions
  (Eq.~\ref{eq:potential_energy_no_longrange}) shows qualitatively
  wrong asymptotic behaviour (see also Fig.~\ref{fig:sn2_errors} for a
  comparison of prediction errors between both models).}
	\label{fig:sn2_mep}
\end{figure}

\begin{figure}[htbp]
\centering
\captionsetup{width=1.0\textwidth}
\includegraphics[width=\textwidth]{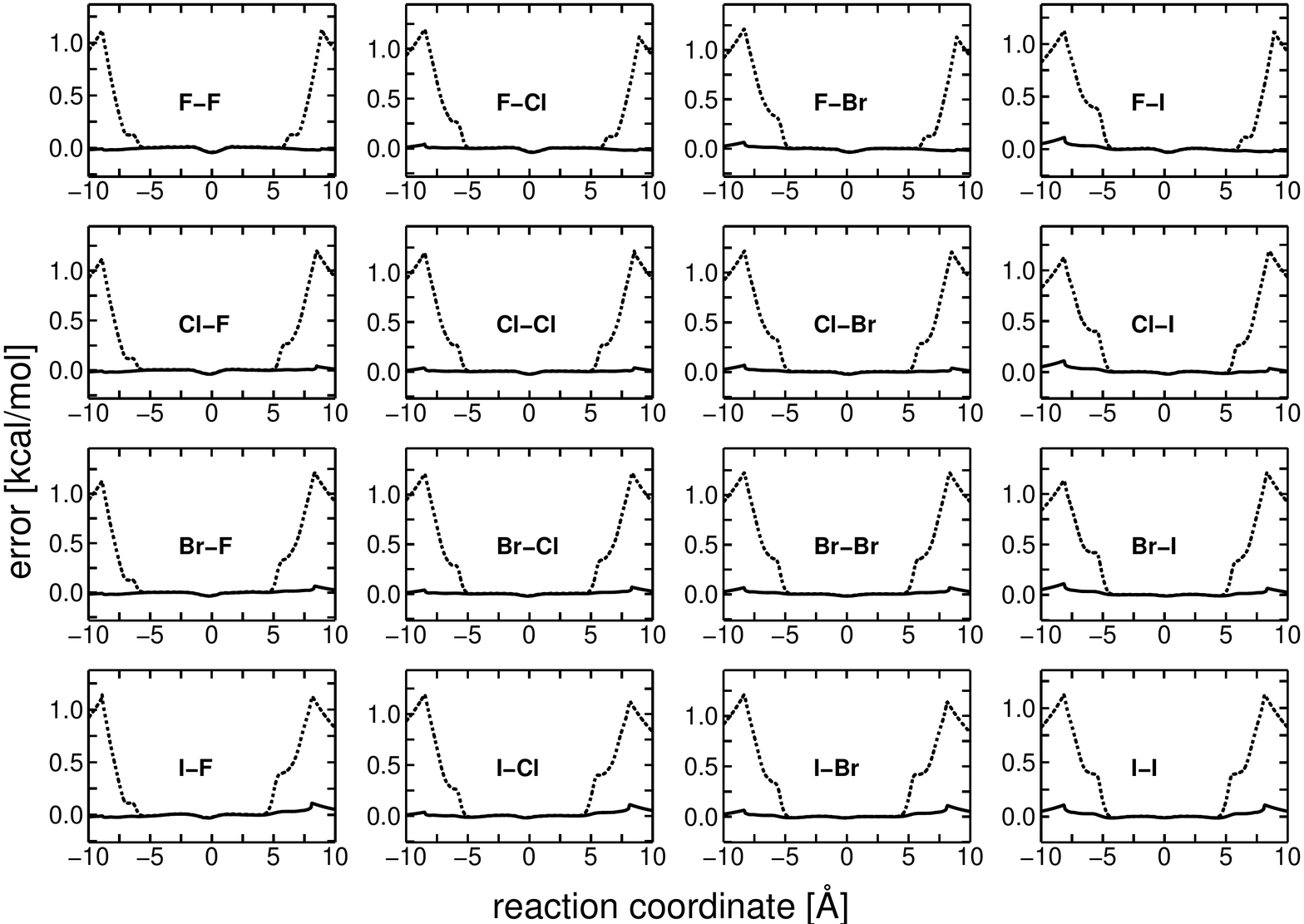}
\caption{Energy prediction errors for minimum energy paths (MEPs) of
  S$_{\rm N}$2 reactions of the kind X$^{-}$ + H$_3$C-Y $\rightarrow$
  X-CH$_3$ + Y$^{-}$ along the reaction coordinate defined by the
  distance difference $r_{\rm CY}-r_{\rm CX}$. The \nn-ens5 models
  with (solid black line) and without (dotted black line) explicit
  long-range interactions are compared. Each panel shows a different
  combination X-Y with X, Y $\in$ \{F, Cl, Br, I\}. The model without
  long-range interactions displays significant errors
  ($\approx1$~kcal~mol$^{-1}$) in the asymptotic regions of the PES.}
	\label{fig:sn2_errors}
\end{figure}

The model without explicit inclusion of long-range interactions
(Eq.~\ref{eq:potential_energy_no_longrange}) performs significantly
worse. This is to be expected, as for this dataset, ion-dipole
interactions, which decay with the square of the distance, play an
important role for determining the overall energy. As their influence
extends well beyond the cut-off distance $r_{\rm cut}$ (here
$10$~\AA), a model without long-range augmentation cannot properly
account for them. This effect is also seen in Fig.~\ref{fig:sn2_mep},
which shows minimum energy paths (MEPs) for all S$_{\rm N}$2 reactions
of the kind X$^{-}$ + H$_3$C--Y $\rightarrow$ X--CH$_3$ + Y$^{-}$
covered in the dataset (all possible combinations X--Y with X,Y $\in$
\{F, Cl, Br, I\}) along the reaction coordinate defined by the
distance difference $r_{\rm CY}-r_{\rm CX}$.  While \nn\ including
explicit long-range interactions
(Eq.~\ref{eq:potential_energy_with_longrange}) is able to reproduce
the reference energies accurately across the whole range of values of
the reaction coordinate, the NN without long-range interactions
(Eq.~\ref{eq:potential_energy_no_longrange}) shows qualitatively wrong
asymptotic behaviour (see
Figs.~\ref{fig:sn2_mep}~and~\ref{fig:sn2_errors}). A correct
description of the asymptotics is crucial for quantitative predictions
of reaction rates with MD simulations, as errors can strongly
influence the maximum impact parameter for collisions at which a
reaction is still possible. Including the long-range behaviour in the
functional form of an ML model has been used previously in the
construction of PESs with kernel ridge regression and has the
additional advantage that less reference data is needed in asymptotic
regions.\cite{unke2017toolkit} Note that the MEPs calculated with the
reference \textit{ab initio} method are not included in the training
data.

\paragraph{Solvated protein fragments}
For a detailed description of the dataset, see
section~\ref{sec:dataset_generation:solvated_protein_fragments}. The
task is to predict all properties (energy, forces and dipole moments)
using a single model, which tests the generalization capabilities
across chemical and conformational degrees of freedom in gas and
solution phase, proton transfer reactions and challenging many-body
intermolecular interactions.

Table~\ref{tab:solvated_protein_fragments_results} lists the
performance of five \nn\ models (including long-range interactions,
see Eq.~\ref{eq:potential_energy_with_longrange}) and their
ensemble\cite{breiman1996bagging} prediction (\nn-ens5). All models
were trained on 2560k structures with 100k structures used for
validation. Because the solvated protein fragments dataset contains
structures with widely different numbers of atoms (up to 120), the MAE
for energy predictions per atom is also reported.

\begin{table}[htbp]
\caption{Mean absolute errors for predictions of energy and energy per
  atom (in kcal mol$^{-1}$), forces (in kcal mol$^{-1}$) \AA$^{-1}$
  and dipole moments (in D) for the solvated protein fragments
  dataset. Results for \nn\ are averaged over five independent runs,
  whereas \nn-ens5 is the performance of an
  ensemble\cite{breiman1996bagging} of five \nn\ models. The best
  results in each category are shown in bold. Note that because of the
  charge correction scheme (see
  Eq.\ref{eq:corrected_partial_charges}), total charge is always
  predicted exactly. However, for completeness, the error for the
  prediction of total charge (in $e$) using the uncorrected partial
  charges is also reported.}
\label{tab:solvated_protein_fragments_results}
\centering
\begin{tabular}{c S S }
		\toprule
		& 
		\multicolumn{1}{c}{\textbf{\nn}} &
		\multicolumn{1}{c}{\textbf{\nn-ens5}}  \\
		\midrule
		\textit{energy} & 1.03 &  \boldentry{1.2}{0.95} \\
		\addlinespace[1pt]
		\textit{energy/atom} & 0.054 & \boldentry{2.3}{0.050}  \\
		\addlinespace[1pt]
		\textit{forces} & 0.88 & \boldentry{1.2}{0.72} \\
		\addlinespace[1pt]
		\textit{dipole} & 0.060 & \boldentry{2.3}{0.054}  \\
		\addlinespace[1pt]
		\textit{charge}  & 0.004 & \boldentry{2.3}{0.003} \\
		\bottomrule
	\end{tabular}
\end{table}

\begin{figure}[htbp]
\centering
\includegraphics[width=0.75\textwidth]{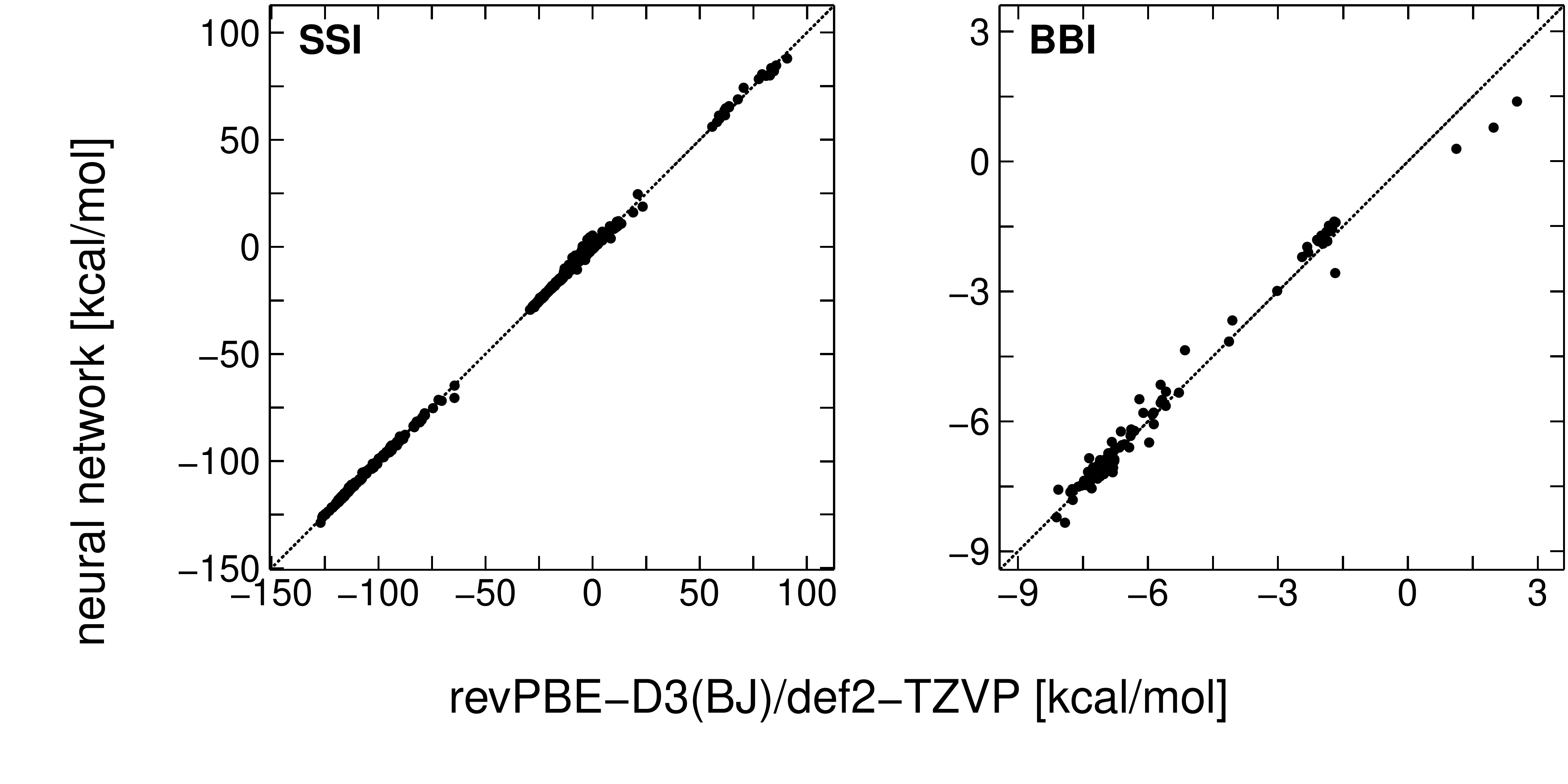}
\caption{Correlation of interaction energies for structures in the
  BioFragment Database\cite{burns2017biofragment} (left: SSI, right:
  BBI) predicted by \nn-ens5 with values obtained from \textit{ab
    initio} calculations (revPBE-D3(BJ)/def2-TZVP). In both cases, the
  predictions correlate well with the reference data (SSI:
  $R^2=0.9997$, BBI: $R^2=0.9922$).}
\label{fig:ssi_bbi_correlation}
\end{figure}

Since non-covalent interactions play a crucial rule for the structure
of large systems like proteins, \nn-ens5 was also used to predict
interaction energies for sidechain-sidechain interactions (SSIs,
\num{3380} structures) and backbone-backbone interactions (BBIs,
\num{100} structures) in the BioFragment
Database\cite{burns2017biofragment} and compared to values calculated
at the reference revPBE-D3(BJ)/def2-TZVP level of theory. For each
case, interaction energies were determined by subtracting monomer
energies from the energy of the complex. The predictions of \nn-ens5
correlate well with the reference values ($R^2>0.99$, see
Fig.~\ref{fig:ssi_bbi_correlation}) and have mean absolute errors of
0.28~kcal~mol$^{-1}$ and 0.21~kcal~mol$^{-1}$ for the SSI and BBI
complexes, respectively. Note that although complexes from the
BioFragment Database are included in the solvated protein fragments
dataset (see
section~\ref{sec:dataset_generation:solvated_protein_fragments}), the
reference data contains only total energies and models were therefore
never directly trained to reproduce interaction energies. Despite this
fact, \nn\ is able to learn a meaningful decomposition of the total
energy into intramolecular and intermolecular contributions and
predict interaction energies accurately.

\begin{figure}[htbp]
\centering
\includegraphics[width=0.75\textwidth]{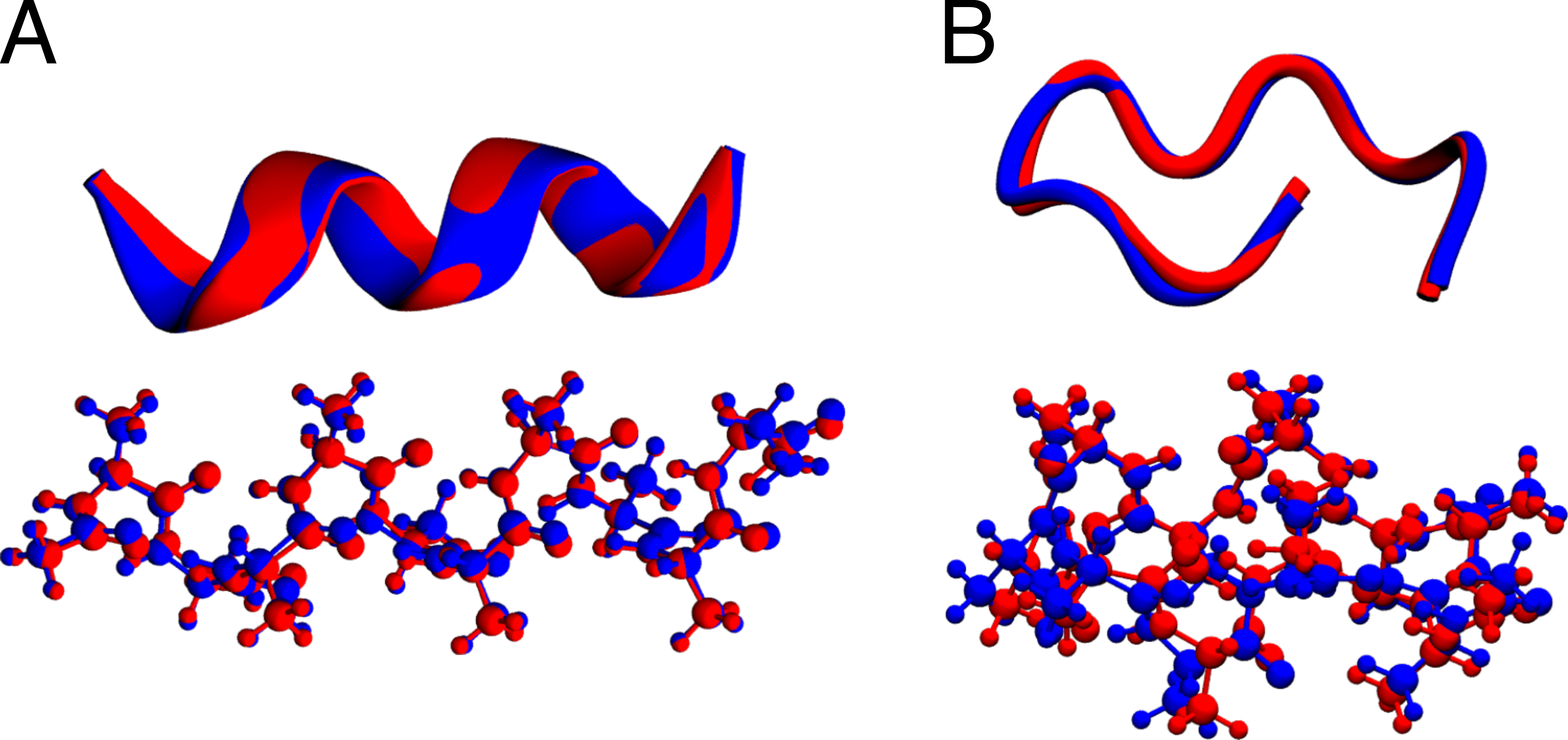}
\caption{Optimized structures of (A) helical Ala$_{10}$ and (B)
  wreath-shaped Ala$_{10}$ in Cartoon representation (top) and as
  ball-and-stick model (bottom). The structures obtained using
  \nn-ens5 (red) and the reference revPBE-D3(BJ)/def2-TZVP method
  (blue) are superimposed in order to highlight differences
  (A:~$\mathrm{RMSD}=0.21$~\AA, B:~$\mathrm{RMSD}=0.52$~\AA).}
\label{fig:decaala}
\end{figure}

In order to test whether predictions also generalize to larger
molecules, deca-alanine (Ala$_{10}$), which is a widely used model
system to study protein folding
dynamics,\cite{hazel2014thermodynamics} is considered as a test
case. Starting from a previously published helical structure of
Ala$_{10}$ (capped with an acetylated N-terminus and amidated
C-terminus),\cite{park2003free} its geometry was optimized with the
BFGS algorithm\cite{fletcher2013practical} using \nn-ens5, as well as
revPBE-D3(BJ)/def2-TZVP to determine the necessary energy
gradients. The energies (relative to free atoms) of the optimized
structures are $-11339.49$ kcal~mol$^{-1}$ and $-11317.05$
kcal~mol$^{-1}$ for the \textit{ab initio} method and \nn-ens5,
respectively, which corresponds to a relative prediction error of
about $0.20$\%. Although \nn-ens5 predicts the optimized structure to
be about $0.207$~kcal~mol$^{-1}$~atom$^{-1}$ less stable than the
\textit{ab initio} method, both optimized geometries are structurally
almost indistinguishable ($\mathrm{RMSD}=0.21$~\AA, see
Fig.~\ref{fig:decaala}A). This result is remarkable, considering that
the ``protein fragments'' in the solvated proteins dataset contain at
most eight heavy atoms (see
section~\ref{sec:dataset_generation:solvated_protein_fragments}),
whereas Ala$_{10}$ consists of \num{54} heavy atoms (\num{109} atoms
in total).

As a final test, the folding of Ala$_{10}$ was investigated by running
unbiased Langevin dynamics\cite{langevin1908theorie} with the
ASE\cite{larsen2017atomic} at a temperature of $300$~K and using a
time step of $0.1$~fs. The necessary forces were obtained from the
predictions of \nn-ens5. Starting from the optimized structure of
stretched Ala$_{10}$, the simulation was run for a total of
\num{400000} time steps ($40$~ps). After about $30$~ps of simulation,
Ala$_{10}$ folds into a wreath-shaped structure (see
Fig.~\ref{fig:decaala}B), in which it remains for the remainder of the
simulation.

\begin{figure}[htbp]
	\centering
	\includegraphics[width=0.75\textwidth]{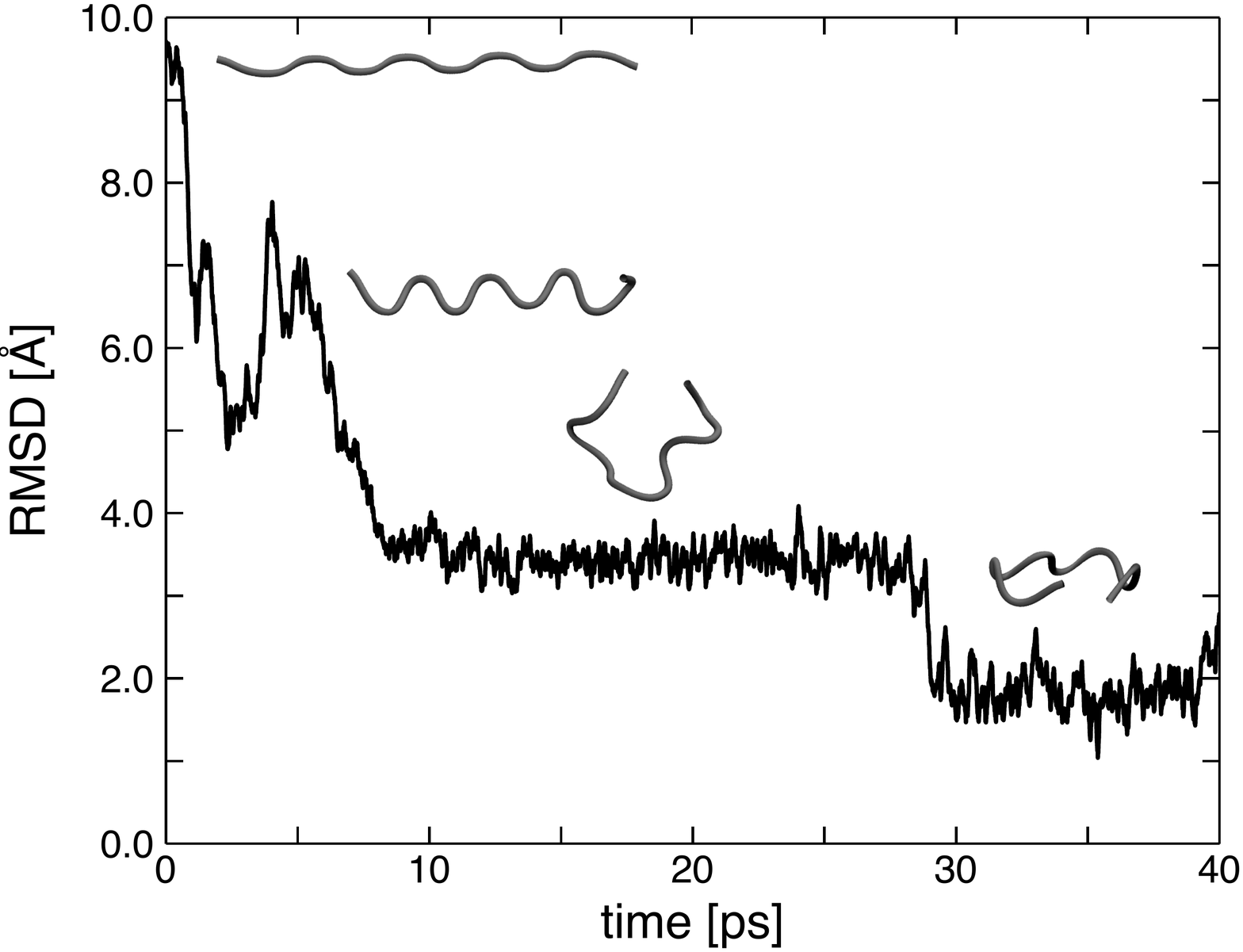}
	\caption{RMSD of Ala$_{10}$ with respect to the optimized
		wreath-shaped geometry (see Fig.~\ref{fig:decaala}B) over the course
		of a \num{40}~ps MD trajectory computed with \nn-ens5. Cartoon
		representations of the structure for representative snapshots along
		the trajectory are shown as well.}
	\label{fig:rmsd_trajectory}
\end{figure}

In order to determine whether the PES explored during the dynamics is
representative of the PES computed using the reference method, the
energy of \num{20} structures sampled at \num{2}~ps intervals along
the trajectory was evaluated with \nn-ens5 and
revPBE-D3(BJ)/def2-TZVP. On average, the prediction error for these
structures is $0.233$~kcal~mol$^{-1}$~atom$^{-1}$ (0.23\% relative
error), with minimum and maximum errors of
$0.072$~kcal~mol$^{-1}$~atom$^{-1}$ (0.07\% relative error) and
$0.405$~kcal~mol$^{-1}$~atom$^{-1}$ (0.39\% relative error),
respectively. Finally, to determine whether the wreath-shaped
conformation obtained at the end of the trajectory is a local minimum
on the Ala$_{10}$ PES, its geometry was optimized with BFGS using
\nn-ens5, as well as revPBE-D3(BJ)/def2-TZVP to determine the
necessary energy gradients. The energies (relative to free atoms) of
the optimized structures are $-11339.95$ kcal~mol$^{-1}$ and
$-11337.07$ kcal~mol$^{-1}$ for the \textit{ab initio} method and the
\nn-ens5, respectively (which corresponds to a relative error of about
$0.03$\%). While both optimized geometries are structurally similar
($\mathrm{RMSD}=0.52$~\AA, see Fig.~\ref{fig:decaala}B), the \nn-ens5
predicts the wreath-shaped geometry to be more stable than the helical
form by about $0.184$~kcal~mol$^{-1}$~atom$^{-1}$, whereas according
to the \textit{ab initio} method, both structures have almost the same
energy (the wreath-shaped geometry is still more stable, but only by
about $0.004$~kcal~mol$^{-1}$~atom$^{-1}$). The RMSD of Ala$_{10}$
with respect to the optimized wreath-shaped structure along the
trajectory is shown in Fig.~\ref{fig:rmsd_trajectory}.

One of the attractive future prospects of ML-based PESs is the
possibility that they become accurate and computationally efficient
alternatives\cite{chmiela2017machine} to empirical force fields (FFs).
Such FFs are parametrized, empirical functions to describe chemical
and biological systems and their parameters are fitted to a
combination of experimental (e.g.\ X-ray structures, infrared
spectroscopy, or hydration free energies) and computer-generated data
(e.g.\ partial charges).\cite{mackerell2004} From this perspective, it
is instructive to consider the question whether an empirical FF yields
similar results as \nn-ens5 when compared to \textit{ab initio}
values. For this, the energy of 20 structures sampled along the
trajectory (see above) was evaluated with the CHARMM
program\cite{brooks1983charmm} using the CHARMM36 all atom
FF.\cite{best2012optimization} In addition, the helical and
wreath-shaped structures were (re)optimized using the FF and compared
to those obtained from \nn-ens5 and revPBE-D3(BJ)/def2-TZVP. For both
FF-optimized structures, the RMSD is $0.29$~\AA\ with respect to the
structures obtained from revPBE-D3(BJ)/def2-TZVP, which is comparable
to RMSDs of $0.21$~\AA\ and $0.52$~\AA\ for the helical and
wreath-shaped structures obtained using \nn-ens5 (see above). To allow
a meaningful comparison between FF and \textit{ab initio} energies,
all values are taken relative to those of the optimized wreath-shaped
structure (the minimum energy according to
revPBE-D3(BJ)/def2-TZVP). On average, the FF energies differ by
$0.430$~kcal~mol$^{-1}$~atom$^{-1}$ from the revPBE-D3(BJ)/def2-TZVP
reference values, which is approximately a factor of $2$ larger than
energies from \nn-ens5 ($0.233$~kcal~mol$^{-1}$~atom$^{-1}$). Given
the faster evaluation time of an empirical FF (by a factor of
$\approx$10--100), this performance appears to be acceptable. It
should be noted that the standard CHARMM36-FF does not contain
refinements such as multipolar
interactions\cite{stone1981distributed,kramer2012atomic} or
polarization effects,\cite{lamoureux2003simple} which can increase the
accuracy of FFs, but also make them less computationally efficient.

% % % % % % % % % % % % % % % % % % % % % % % % % % % % % % %
% DISCUSSION AND CONCLUSION
% % % % % % % % % % % % % % % % % % % % % % % % % % % % % % %
\section{Discussion and Conclusion}
\label{sec:discussion_and_conclusion}
In the present work, the \nn\ neural network architecture was
introduced and tested on several common quantum-chemical benchmark
datasets.  It matches or improves state-of-the-art performance of
machine learning models for all tested benchmarks, in some cases
decreasing the error of previously published models between 50--90\%.
While kernel-based approaches like GDML\cite{chmiela2017machine} (or
its successor sGDML\cite{chmiela2018towards}) achieve similar or
sometimes even better performance on some benchmarks like MD17,
kernel-based methods have the disadvantage that the computational cost
of evaluating them scales linearly with training set size. The cost of
evaluating \nn\ on the other hand is independent of the amount of data
used for training and only depends on the chosen architecture. For
this reason, neural network based models are better suited for
constructing PESs that are transferable between many different
chemical systems and require large amounts of training data (see for
example the solvated protein fragments benchmark).

Two new datasets were introduced that address chemical situations not
covered in other published datasets, namely chemical reactivity and
many-body intermolecular interactions important for condensed phase
systems. It was demonstrated that incorporating physical knowledge
into the model (e.g.\ electrostatic contributions) can be crucial for
its predictive quality.

While optimized helical structures of Ala$_{10}$ using the \nn-PES
(trained on the solvated protein fragments dataset) and the reference
\textit{ab initio} method are almost identical (see
Fig.~\ref{fig:decaala}A), the relative error in the energy prediction
of \nn-ens5 is about an order of magnitude larger than for the
wreath-shaped structure of Ala$_{10}$ (see
Figure~\ref{fig:decaala}B). A possible explanation for this
discrepancy could be the large dipole moment of helical protein
structures due to the cumulative effect of the individual dipole
moments of carbonyl groups aligned along the helix
axis.\cite{hol1978alpha} The electric field associated with a large
dipole moment likely leads to strong polarization effects, which
potentially influence the total energy substantially. While
polarization effects can be captured implicitly by \nn\ due to its
ability to assign environment-dependent partial charges to atoms, it
is likely that the structures included in the training data do not
contain sufficient information to describe the cumulative polarization
effects of multiple aligned dipole moments. A larger dataset of
reference structures including helical motifs would likely be needed
for a proper description of such phenomena.

In summary, \nn\ is able to accurately predict energies and forces for
a wide range of structures across chemical and conformational degrees
of freedom and different datasets. For S$_{\rm N}$2 reactions of
methyl halides with halide anions, it was shown that including
long-range electrostatic interactions explicitly in the model
significantly improves the qualitative shape of the predicted PES
close to and beyond the cut-off radius. Further, it was shown that
\nn\ can distinguish between intra- and intermolecular contributions
in SSIs and BBIs of proteins in a meaningful manner. When trained on a
large set of small reference structures, the \nn\ model is able to
generalize to larger structures like Ala$_{10}$ with similar
structural motifs. This result suggests that with a systematically
constructed set of small reference structures, it is possible to build
a transferable model applicable to a wide range of chemical
systems. However, some large-scale effects, for example strong
electric fields due to multiple aligned microscopic dipole moments,
might not be properly accounted for when training only on small
molecules.

% % % % % % % % % % % % % % % % % % % % % % % % % % % % % % %
% ACKNOWLEDGEMENTS
% % % % % % % % % % % % % % % % % % % % % % % % % % % % % % %
\section*{Acknowledgments}
\label{sec:acknowledgements}
The authors acknowledge financial support from the Swiss National
Science Foundation (NCCR-MUST and Grant No. 200021-7117810) and the
University of Basel.

\section*{Dataset and code availability}
The S$_{\rm N}$2 reactions\cite{sn2_reactions_dataset} and the solvated protein fragments\cite{solvated_protein_fragments_dataset} datasets are available from \url{www.quantum-machine.org} or directly from [\citenum{sn2_reactions_dataset,solvated_protein_fragments_dataset}]. The code for training \nn\ models is freely available from \url{https://github.com/MeuwlyGroup/PhysNet}.

% % % % % % % % % % % % % % % % % % % % % % % % % % % % % % %
% SUPPORTING INFORMATION
% % % % % % % % % % % % % % % % % % % % % % % % % % % % % % %

%\section*{Supporting Information}
%\label{sec:supporting_information}
%The supporting information to this article contains details on the generation of the datasets introduced in this work and additional figures.

\bibliography{references}

\end{document}